\documentclass[11pt,a4paper]{emulateapj}

\usepackage{psfrag}
\usepackage{psfig}
\usepackage{pspicture}
\usepackage{graphicx}

\usepackage{amssymb,amsmath}
\usepackage{natbib}

\def\gs{\mathrel{\raise0.35ex\hbox{$\scriptstyle >$}\kern-0.6em\lower0.40ex\hbox{{$\scriptstyle \sim$}}}} 
\def\ls{\mathrel{\raise0.35ex\hbox{$\scriptstyle <$}\kern-0.6em\lower0.40ex\hbox{{$\scriptstyle \sim$}}}} 
\def\Msol{\mathrel{\rm M_{\odot}}} 
\def\Msolyr{\mathrel{\rm M_{\odot}\,yr^{-1}}} 
\def\gsim{\mathrel{\raise0.35ex\hbox{$\scriptstyle >$}\kern-0.6em\lower0.40ex\hbox{{$\scriptstyle \sim$}}}} 
\def\lsim{\mathrel{\raise0.35ex\hbox{$\scriptstyle <$}\kern-0.6em\lower0.40ex\hbox{{$\scriptstyle \sim$}}}} 
\def\ltsima{$\; \buildrel < \over \sim \;$} 
\def\simlt{\lower.5ex\hbox{\ltsima}} 
\def\gtsima{$\; \buildrel > \over \sim \;$} 
\def\simgt{\lower.5ex\hbox{\gtsima}}

\begin{document}

\title {Cosmic Web and Star Formation Activity in Galaxies at \lowercase{$z\sim$}1}

\author{
B.\ Darvish,\altaffilmark{1}
D.\ Sobral,\altaffilmark{2,3,4}
B.\, Mobasher,\altaffilmark{1}
N.\, Z.\, Scoville,\altaffilmark{5}
P.\, Best,\altaffilmark{6}
L.\, V.\, Sales,\altaffilmark{1,7}
I.\, Smail\altaffilmark{8}
}
\setcounter{footnote}{0}
\altaffiltext{1}{University of California, Riverside, 900 University Ave, Riverside, CA, 92521, USA; email: bdarv001@ucr.edu}
\altaffiltext{2}{Instituto de Astrof\'{\i}sica e Ci\^encias do Espa\c co, Universidade de Lisboa, OAL, Tapada da Ajuda, PT 1349-018 Lisboa, Portugal}
\altaffiltext{3}{Centro de Astronomia e  Astrof\'{\i}sica da Universidade de Lisboa, Observat\'{o}rio Astron\'{o}mico de Lisboa, Tapada da Ajuda, 1349-018 Lisboa, Portugal}
\altaffiltext{4}{Leiden Observatory, Leiden University, P.O. Box 9513, NL-2300 RA Leiden, The Netherlands}
\altaffiltext{5}{California Institute of Technology, MC 249-17, 1200 East California Boulevard, Pasadena, CA 91125, USA}
\altaffiltext{6}{SUPA, Institute for Astronomy, Royal Observatory of Edinburgh, Blackford Hill, Edinburgh, EH9 3HJ, UK}
\altaffiltext{7}{Harvard-Smithsonian Center for Astrophysics, 60 Garden Street, Cambridge, MA 02138, USA}
\altaffiltext{8}{Institute for Computational Cosmology, Durham University, South Road, Durham DH1 3LE, UK}
\begin{abstract}
We investigate the role of the delineated cosmic web/filaments on the star formation activity by exploring a sample of 425 narrow-band selected H$\alpha$ emitters, as well as 2846 color-color selected underlying star-forming galaxies for a large scale structure (LSS) at z=0.84 in the COSMOS field from the HiZELS survey. Using the scale-independent Multi-scale Morphology Filter (MMF) algorithm, we are able to quantitatively describe the density field and disentangle it into its major components: fields, filaments and clusters. We show that the observed median star formation rate (SFR), stellar mass, specific star formation rate (sSFR), the mean SFR-Mass relation and its scatter for both H$\alpha$ emitters and underlying star-forming galaxies do not strongly depend on different classes of environment, in agreement with previous studies. However, the fraction of H$\alpha$ emitters varies with environment and is enhanced in filamentary structures at z$\sim$1. We propose mild galaxy-galaxy interactions as the possible physical agent for the elevation of the fraction of H$\alpha$ star-forming galaxies in filaments. 
Our results show that filaments are the likely physical environments which are often classed as the ``intermediate" densities, and that the cosmic web likely plays a major role in galaxy formation and evolution which has so far been poorly investigated.     

\end{abstract}


\section{Introduction} \label{intro}
It is well known that the properties of galaxies are directly affected by their host environment. In the local Universe, red, passive, early-type galaxies reign over-densities and galaxy clusters whilst blue, star-forming, late-type galaxies are preferentially found in less-dense, field environments \citep[e.g.][]{Dressler80,Kauffmann04,Balogh04}. The same trends are seen with stellar mass; i.e., on average, more massive galaxies are redder \citep[e.g.][]{Baldry06}, less star-forming \citep[e.g.][]{Peng10} and more likely have early-type morphologies \citep[e.g.][]{Bamford09}. Since both environment and stellar mass influence the observable properties of galaxies, it is important to consider the effects of both and their fractional role on the evolution of galaxies.\\
For example, a tight relation has been recently found between the SFR and stellar mass of star-forming galaxies; i.e., the ``Main Sequence", showing that more massive star-forming galaxies are more actively forming stars (\citealp{Brinchmann04,Daddi07,Elbaz07,Noeske07a,Karim11,Reddy12} see also \citealp{Speagle14} for a full list of references). The role of the environment on this SFR-Mass relation, along with its slope, intercept and intrinsic scatter, and the evolution of all these parameters with look-back time can impose strong constraints on scenarios of formation and evolution of galaxies \citep{Noeske07b,Daddi07,Whitaker12}. As a result, an increasing number of studies are now focused on the role of environment on the SFR-Mass relation, with rather controversial results especially at z$\gtrsim$1 \citep[e.g.][]{Koyama13a,Zeimann13}.\\
Current studies of the environmental dependence of the SFR in galaxies at z$\gtrsim$1 have led to conflicting results. Some agree with the relations observed in the local Universe \citep{Patel09}, some report no/weak dependence on the environment \citep{Grutzbauch11,Scoville13} and there are even claims for a reversal of the SF-density relation \citep[higher star formation in denser regions][]{Elbaz07,Cooper08,Tran10}. \citealp{Sobral11} also showed that at z$\sim$1, there is an increase of SF activity for star-forming galaxies at intermediate densities, likely associated with galaxy groups, followed by a decline for the richest clusters - and that this is the reason for discrepancies: some studies only reach up to poor group environments, while others only probe rich clusters. \citealp{Sobral11} showed that once the full range of environments is probed and correctly labelled, all the apparently contradictory results can be fully reconciled.\\
Studies of the effect of large scale structure on properties of galaxies have so far been mostly confined to field vs. cluster environments. However, there are intermediate environments such as galaxy groups, outskirts of clusters and filaments which are equally important (e.g. see \citealp{Kodama01}). Thus, a rising number of studies now highlight the importance of intermediate environments, revealing an enhancement of mean SFR \citep{Porter07,Porter08,Sobral11,Coppin12} or the fraction of star-forming galaxies (see section \ref{fraction} for more references) in medium environments. However, the environment in these studies is often loosely defined as the excess number of galaxies or visual inspection of overdensities. Indeed, a density-based definition of environment cannot differentiate between regions with intermediate densities such as filaments, groups and periphery of clusters. These could have the same local number density but host different physical processes (e.g. \citealp{Aragon-Calvo10} showed that a pure density-based structural identification does not provide an accurate description of reality given the overlap among different parts of the LSS and that environment defined only in terms of density fails to incorporate the intrinsic dynamics of the LSS).\\
The explicit role of the geometry and dynamics of the ``cosmic web" \citep{Bond96}---a web-like structure containing dense clusters, sparsely populated voids, planar walls and thread-like filamentary structures linking overdense regions--- in the formation and evolution of galaxies is relatively unexplored. The geometry and dynamics of the cosmic web, as revealed by numerical simulations such as Millennium simulation \citep{Springel05,Boylan-Kolchin09} and large spectroscopic surveys in the local Universe such as 2dFGRS \citep{Colless01} and SDSS \citep{York00}, can be approximated by collapse of a primordial gas cloud along its three main axes through successive stages, first leading to the formation of a wall, subsequently a filament and finally a cluster \citep{Zel'dovich70,Shandarin89}. The difficulty of this natural approach toward the definition of environment lies in the limitation to numerical simulations \citep{Colberg05,Stoica05,Aragon-Calvo07} or spectroscopic surveys in the local Universe \citep{Tempel14}.\\
In this paper, we study the star formation activity in a super-structure ($\sim$10$\times$15 Mpc) at z$\sim$0.8-0.9 in the Cosmic Evolution Survey (COSMOS) field \citep{Scoville07} by identifying a striking filamentary structure linking clusters and groups, which is also traced by the distribution of narrow-band selected star-forming galaxies at this redshift \citep{Sobral11,Sobral13}. Here, the cosmic web is robustly described by the Multi-scale Morphology Filter (MMF) algorithm (section \ref{MMF}) in a consistent and homogeneous manner. H$\alpha$ emission line is used as diagnostic of SF activity. The multi-waveband capabilities, as well as the wealth of ancillary information in the COSMOS field including accurate photometric redshifts (photo-z) ($\Delta$z$\sim$0.01 out to z$\sim$1) and stellar masses ($\Delta$M$\sim$0.1 dex) also equip us with a reliable distribution of underlying control sample of galaxies at z$\sim$0.85.\\
The format of this paper is as follows. In section \ref{data}, we briefly review the data, explain stellar mass and star formation rate measurements and discuss completeness and contamination in the data. Section \ref{lss} presents the methods used to identify the large scale structure. The main results of this paper, along with comparison with other studies are given in section \ref{result} and discussed in section \ref{dis}. We give a summary of this work in section \ref{concl}.\\ 
Throughout this work, we assume a flat concordance $\Lambda$CDM cosmology with H$_{0}$=70 kms$^{-1}$ Mpc$^{-1}$, $\Omega_{m}$=0.3 and $\Omega_{\Lambda}$=0.7. All magnitudes are expressed in the AB system and stellar masses and star formation rates are given assuming a Chabrier IMF.

\section{The Data} \label{data}
For this study, we use a sample of narrow-band selected star-forming galaxies and an underlying control sample (star-forming and quiescent) selected to the same magnitude and mass limits and with the same areal coverage. The star-forming sample is based on H$\alpha$ emission line which directly originates from HII regions in star-forming galaxies and is an excellent tracer of the SFR. The star-forming sample is mainly used as a robust indicator of star formation activity for which environmental effects will be studied. The control sample here is primarily used to identify the LSS (section \ref{lss}) and to determine the fraction of H$\alpha$ star-forming galaxies associated with the LSS (section \ref{fraction}). The star-forming galaxies in the control sample are also used to double-check the validity of results based on the H$\alpha$ selected star-forming sample and to investigate the role of different SFR estimators on the results. In the following subsections, we describe each of these samples in more detail, along with their stellar mass and SFR estimations and their corresponding completeness and contamination. 
     
\subsection{The Star-forming Sample} \label{ha}
This consists of a complete sample of H$\alpha$ emitting star-forming galaxies at z=0.845, selected through narrow-band near-infrared (J filter) observations, performed as a part of the High-z Emission Line Survey (HiZELS) \citep{Geach08,Sobral09,Sobral12,Sobral13}, covering an area of $\sim$0.8 deg$^{2}$ in the COSMOS field \citep{Scoville07}. The sample includes a total of 425 galaxies with K$<$23 and log(M/$\Msol)\geq$9, located in a narrow redshift slice of $\Delta$z=0.03 centered at z=0.845 (0.83$\leq$z$\leq$0.86). This sample is flux limited ($\sim 8\times 10^{-17}$ ergs$^{-1}$cm$^{-2}$ which is equivalent to the SFR limit of $\sim$ 1.5 $\Msolyr$ at z$\sim$0.84) and is selected homogeneously down to the rest-frame H$\alpha$+[NII] equivalent width limit of $\sim$ 25 {\AA}. For more properties of this sample, see e.g. \citealp{Sobral11} \& \citealp{Sobral14}.

\subsection{The Control Sample} \label{control}
\begin{figure}
  \centering
  \includegraphics[width=3.5in]{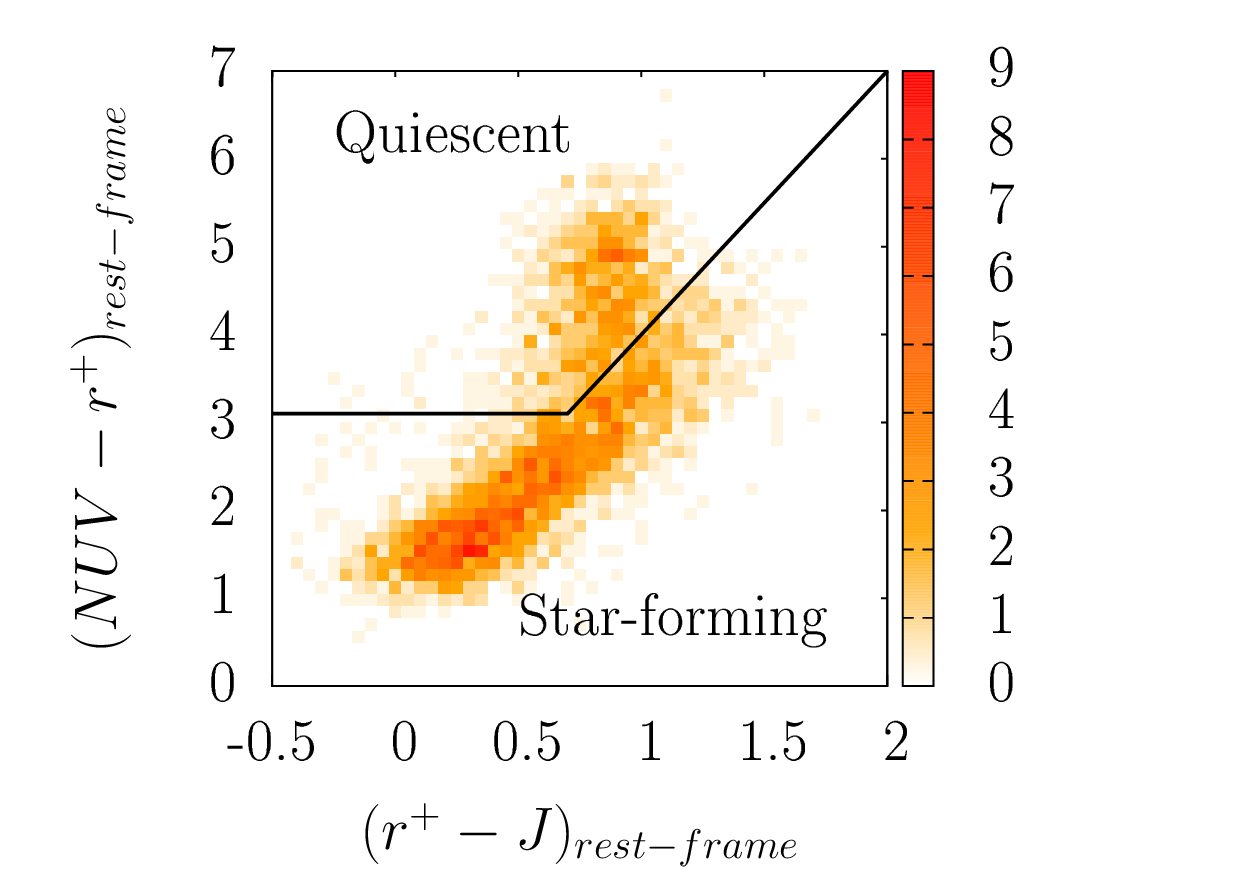}
\caption{ Rest-frame NUV-$r^{+}$ versus $r^{+}$-J color distribution of galaxies in the control sample. Two populations of galaxies are seen here. The black solid line shows the cut we used to separate the quiescent galaxies from the star-forming systems.}
\label{fig:nuv-r-j}
\end{figure}

The control sample is based on the COSMOS UltraVISTA Ks-band selected photometric redshift catalog \citep{Ilbert13,McCracken12}. The Ks-band selected catalog consists of ground and space based imaging data in 30 bands. Due to the large number of filters covering a broad wavelength range, the photometric redshifts are very secure. It has been shown that at z$<$1.5, the photo-z accuracy is better than 1\% with less than 1\% of catastrophic failure \citep{Ilbert13}. At the redshift of this study, the median redshift uncertainty is $\Delta$z$\sim$0.01. Galaxies in the control sample have the same magnitude, mass and areal coverage as the H$\alpha$ sample and have more than 10\% probability of belonging (P$_{limit}$=0.1) to the redshift slice of H$\alpha$ emitters (0.83$\leq$z$\leq$0.86). This probability weight is calculated by measuring what percentage of the photo-z probability distribution function (PDF) for each galaxy lies within the redshift slice defined by the narrow-band H$\alpha$ sample. This results in a Ks-band selected sample containing 3920 galaxies. Increasing the P$_{limit}$ results in a control sample less contaminated by foreground and background sources. However, this also has an increasing risk of losing sources; i.e., a less complete sample at z$\sim$0.84 (see section \ref{comp-cont}). Here, we use a 10\% cut as a trade-off between contamination and completeness. Nevertheless, we scrutinized a variety of probability limits (0.05, 0.2, 0.3, 0.4 and 0.5) and were able to retrieve the same results.\\ 
The underlying galaxies were later divided into quiescent and star-forming systems based on their rest-frame two-color NUV-$r^{+}$ versus $r^{+}$-J plot. Previous studies show that the rest-frame NUV-$r^{+}$ color is a better indicator of the recent star formation activity \citep{Martin07} and has a wider dynamical range compared to the rest-frame U-V color \citep{Ilbert13}. Here, galaxies with their rest-frame color NUV-$r^{+}$ $>$ 3.1 and NUV-$r^{+}$ $>$ 3($r^{+}$-J)+1 are selected as quiescent systems. Figure \ref{fig:nuv-r-j} shows the rest-frame  NUV-$r^{+}$ versus $r^{+}$-J distribution of galaxies in the control sample and the color cuts used to separate quiescent and star-forming systems. We clearly see two populations of galaxies. We stress that adding dust to the star-forming galaxies causes them to move diagonally from the bottom left to the top right of Figure \ref{fig:nuv-r-j}, making them separable from the quiescent systems \citep{Ilbert13}. The star-forming galaxies in the control sample will later be used, along with the H$\alpha$ sample, to study the effects of different SFR estimators on the results.
\subsection{Mass \& SFR Estimation} \label{M-SFR}
\subsubsection{Mass \& SFR for the Star-forming Sample} \label{M-SFR-Ha}
For the H$\alpha$ sample, the stellar masses were estimated by SED template fitting using rest-frame UV, optical, near- and mid-IR data available in 18 bands as explained in \citealp{Sobral11} \& \citealp{Sobral14}. The synthetic templates were generated using \citealp{Bruzual07} models, assuming a Chabrier IMF, five different metallicities, an exponentially declining SFH with different e-folding time scales ($\tau$=0.1-10 Gyr) and \citealp{Calzetti00} dust extinction. Stellar masses here are based on the median of the stellar mass PDF, marginalized over all other parameters. The H$\alpha$ sample is complete to log(M/$\Msol)\sim$9-9.5 at z=0.84 and the stellar mass has a typical observational uncertainty of $\Delta$M$\sim$0.1 dex.\\
H$\alpha$ fluxes are obtained by computing the emission line flux within the narrow-band J filter, after removing the contribution from the [NII] line. The [NII] contribution is estimated using the relation between N[II]/H$\alpha$ metallicity and the rest-frame H$\alpha$+[NII] equivalent width from SDSS \citep{Villar08}. This relation seems to hold true at the redshift of our study \citep{Stott13}. H$\alpha$ luminosities are based on 2$^{''}$ diameter aperture photometry ($\sim$16 Kpc physical diameter at z$\sim$0.84). Since we miss $\sim$23\% of the total flux in the 2$^{''}$ aperture \citep{Sobral14}, a correction factor of 1.3 is applied to the measured luminosities. Dust-corrected H$\alpha$ star formation rates were obtained using aperture corrected H$\alpha$ luminosities and the relation from \citealp{Kennicutt98}, modified for a Chabrier IMF: $SFR(\Msolyr)=4.4\times10^{-42}L_{H\alpha}$(ergs$^{-1}$). Dust correction was performed based on the empirical relation between median stellar mass and median dust extinction presented in \citealp{Garn10}. This relation is shown to be applicable to star-forming galaxies at least up to z$\sim$1.5 without evolution (\citealp{Sobral12}; later confirmed by \citealp{Ibar13}, \citealp{Dominguez13} \& \citealp{Price14}). The typical observational uncertainty in the SFR is dominated by the uncertainty in the dust correction ($\Delta$SFR$\sim$0.2 dex). 

\subsubsection{Mass \& SFR for the Control Sample} \label{M-SFR-control}
Stellar masses for galaxies in the control sample were derived as explained in \citealp{Ilbert13}. This was performed by generating a library of synthetic spectra using BC03 \citep{Bruzual03} with a Chabrier IMF, three different metallicities, an exponentially declining SFH with different e-folding time scales ($\tau$=0.1-30 Gyr) and \citealp{Calzetti00} extinction law. Contributions from nebular emission lines were included. These synthetic spectra were fitted to the rest-frame multi-color photometry from UV to mid-IR using ``Le Phare" code \citep{Arnouts02,Ilbert06}. The stellar mass for the control sample corresponds to the median of the stellar mass PDF, marginalized over all other parameters. At z$\sim$1, we detect galaxies with masses as low as log(M/$\Msol)\sim$8.9 given a sample selected at Ks$<$23. We estimate the mass completeness limit following the procedure explained in \citealp{Ilbert13} \& \citealp{Pozzetti10}. The typical observational uncertainty of stellar mass for the control sample is $\Delta$M$\sim$0.1 dex. These stellar masses are also available for the majority of H$\alpha$ emitters.\\
Dust-corrected star formation rates are derived from the rest frame NUV continuum (NUV corresponds to the GALEX filter centered on 0.23$\mu$m) and the Spitzer 24$\mu$m flux (for galaxies with 24$\mu$m detections). Rest-frame observed and extinction-corrected NUV flux were calculated in \citealp{Ilbert13}. GALEX NUV flux values were estimated from the BC03 best-fit photo-z SED templates using the NUV magnitudes computed by \citealp{Zamojski07}. The Spitzer 24$\mu$m flux is extracted from the available catalogs as part of the S-COSMOS survey \citep{Sanders07}. MIPS 24$\mu$m total flux was computed by Aussel and Le Floc'h \citep{LeFloch09} assuming a 10000K blackbody as the underlying spectrum. The SFRs are estimated in one of the two ways explained below:\\
\begin{itemize}
\item For galaxies with no 24$\mu$m detection, we make use of the equation given in \citealp{Kennicutt98}, modified for the Chabrier IMF, in order to estimate SFR based on the dust-corrected NUV continuum flux ($L_{NUV}^{corr}$): $SFR(\Msolyr)=0.786\times10^{-28}L_{NUV}^{corr}(ergs^{-1}Hz^{-1})$. 
\item For galaxies with available 24$\mu$m flux, the SFR is estimated based on the observed 24$\mu$m flux density using the relation given in \citealp{Rieke09}: $log(SFR(\Msolyr))=A(z)+B(z)(log(4\pi D_{L}^2f_{obs})-53)$, where A(z) and B(z) are redshift and wavelength dependent constants, $D_{L}$ is the luminosity distance in cm and $f_{obs}$ is the observed 24$\mu$m flux density in Jy. Here, we use A$_{24}$(z=0.8)=0.445 and B$_{24}$(z=0.8)=1.381 \citep{Rieke09}.
\end{itemize}
The typical uncertainty in the SFR for the control sample is $\Delta$SFR$\sim$0.1 dex.\\

\subsection{Completeness and Contamination} \label{comp-cont}
\begin{figure}
  \centering
  \includegraphics[width=3.5in]{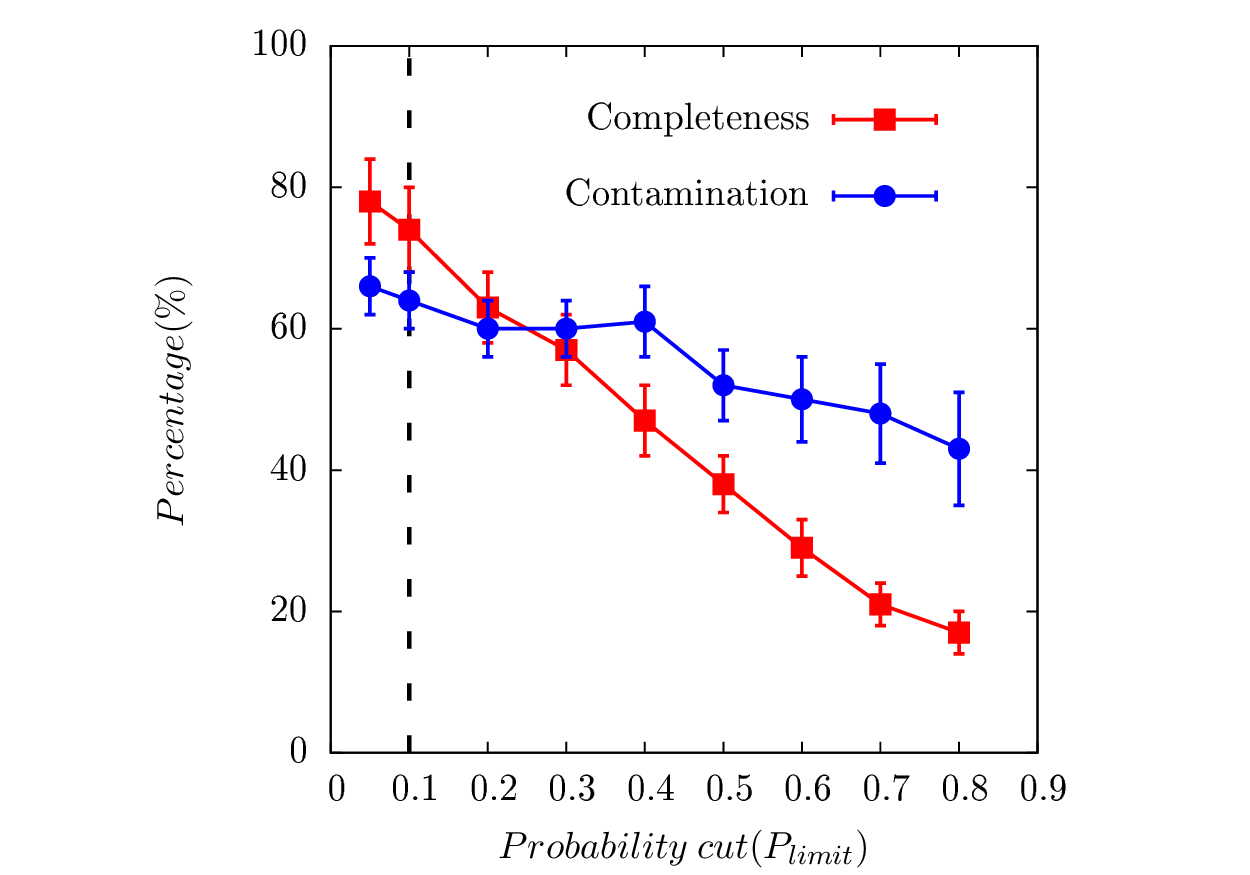}
\caption{ Completeness and contamination functions for the control sample as a function of the probability limit, P$_{limit}$. The probability limit is defined as the minimum probability of galaxies in the control sample to belong to the redshift slice defined by the H$\alpha$ narrow-band selected sample (0.83$\leq$z$\leq$0.86). Error bars are estimated assuming Poisson statistics. We estimate the completeness by counting the number of underlying galaxies in the control sample that have spectroscopic redshifts within the H$\alpha$ redshift slice, compared to the total number of galaxies with spectroscopic redshifts in the same slice. Contamination is defined as the fraction of galaxies in the underlying sample with spectroscopic redshifts located outside the H$\alpha$ redshift slice. Choosing a larger P$_{limit}$ results in a less contaminated but simultaneously less complete and statistically smaller control sample. The control sample here is selected to have a probability limit P$_{limit}$=0.1, demonstrated by the dashed line in the figure.}
\label{fig:comp-cont}
\end{figure}

In order to estimate the completeness and contamination in both the H$\alpha$ and the underlying control samples, we use the VLT-VIMOS zCOSMOS spectroscopic survey \citep{Lilly07} selected as I$<$22.5. Here, only the highly reliable spectroscopic redshifts are considered (Classes 3 \& 4 with spectroscopic reliability $>$ 99.5\%). Using the spectroscopic data, we estimate the completeness as the ratio of the number of underlying galaxies in the control sample that have spectroscopic redshifts within the H$\alpha$ redshift slice (0.83$\leq$z$\leq$0.86) to the total number of galaxies with spectroscopic redshifts in the same slice. Contamination is computed as the fraction of galaxies in the underlying control sample with spectroscopic redshifts located outside the H$\alpha$ redshift slice. Previous studies show the H$\alpha$ sample to be $<$ 4\% contaminated and $>$ 96\% complete (see \citealp{Sobral10} \& \citealp{Sobral13}). For the control sample, we estimate the contamination and completeness to be 64\% and 74\%, respectively (Figure \ref{fig:comp-cont}). As explained in section \ref{M-SFR-control}, a probability limit of 10\% (P$_{limit}$=0.1) was used to select galaxies in the control sample; i.e., galaxies have a likelihood of $>$ 10\% belonging to the redshift slice 0.83$\leq$z$\leq$0.86 set by the H$\alpha$ sample. A higher probability limit results in a more reliable and less contaminated control sample. However, this also leads to a less complete and statistically smaller sample. Figure \ref{fig:comp-cont} shows completeness and contamination fractions in the control sample as a function of the probability limit (P$_{limit}$) used to define it. Completeness declines with increasing probability limit faster than contamination decreases.\\
In estimating the contamination and completeness for the control sample, we assume that they are independent of the apparent magnitude and we can extrapolate the results from bright (I$<$22.5) to faint galaxies. This is a valid assumption given that (1) at I$<$24, a comparison between the  photometric and spectroscopic data shows that the photo-z accuracy is $\sim$0.01 \citep{Ilbert09}, smaller than the redshift slice defined by the H$\alpha$ sample ($\sim$0.03) and (2) the majority of galaxies in the control sample ($\sim$93\%) are brighter than I$<$24.\\
We also stress that since massive red galaxies which dominate denser environments will have sharply peaked photo-z PDFs compared to less massive blue systems, they will be more complete and less contaminated. As a result, the completeness/contamination might be a function of the environment. We investigate the environmental dependence of completeness/contamination and find that although the completeness and contamination functions vary with the environment (highest completeness/lowest contamination in clusters), the correction factor (used in sections \ref{kernel} \& \ref{fraction}) is almost independent of the environment (the higher completeness is accompanied by a lower contamination which leaves the correction factor almost intact) and this will not affect our results.        

\section{Large Scale Structure Identification} \label{lss}
In this section, we use the underlying distribution of galaxies in the control sample to estimate 
the local surface density field of galaxies from which the cosmic web (filaments \& clusters) is extracted.
\begin{figure*}
\centering
\includegraphics[width=7.0in]{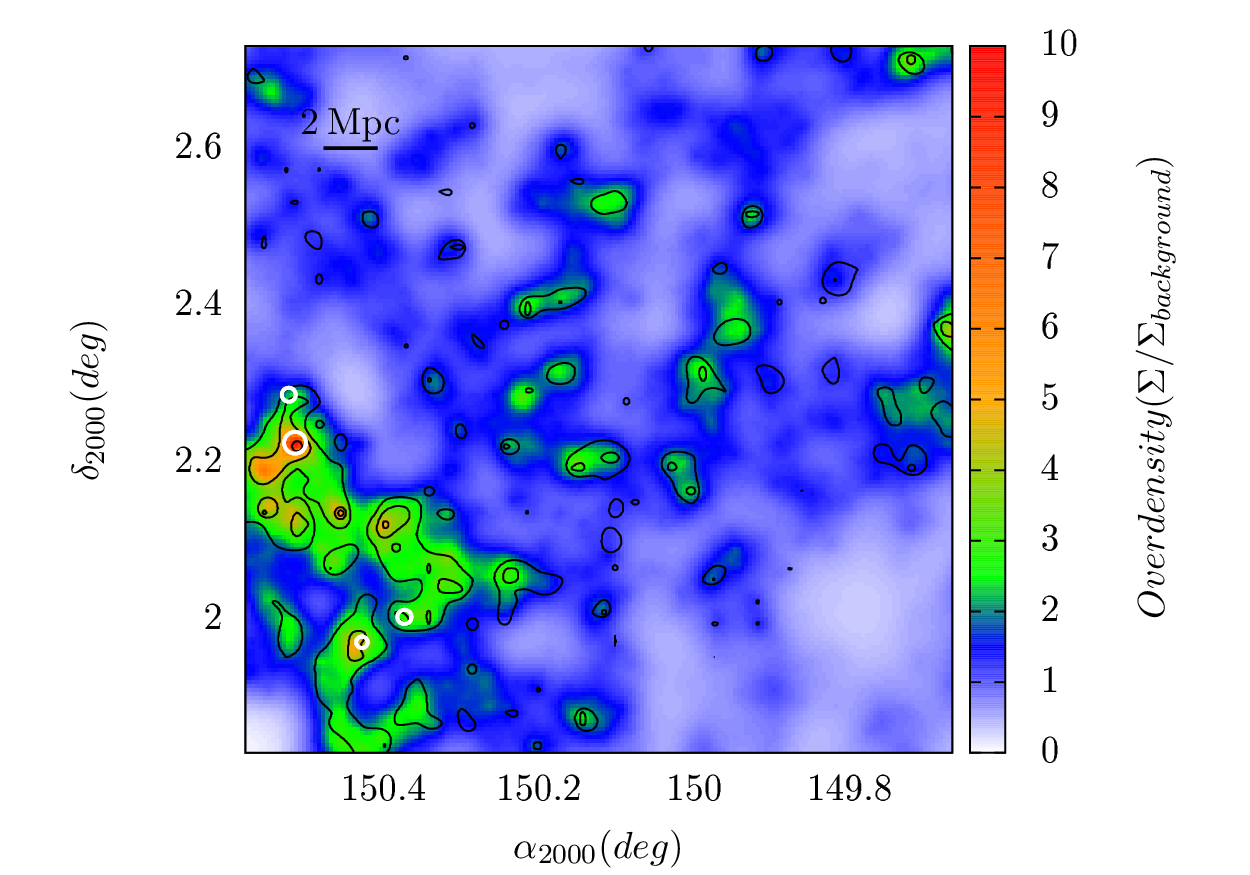}
\caption{ The overdensity map constructed from the Ks-band selected control sample, using the weighted adaptive kernel density estimator for the redshift slice 0.83$\leq$z$\leq$0.86 in the HiZELS area of the COSMOS field. Overdensity is defined as the surface density normalized to the mean surface density at that redshift. Black contours are extracted from \citealp{Scoville13} voronoi tessellation density estimator for the redshift slice 0.83$\lesssim$z$\lesssim$0.85. For clarity, only three contour levels are demonstrated which are at 1/2, 1/4 and 1/8th of the overdensity peak. White circles also show the $R_{200}$ region of the X-ray clusters/groups in redshift range 0.82$\leq$z$\leq$0.87 over the same area. There is a very good consistency between our density estimation and that of \citealp{Scoville13}. Note the super-structure in the lower left side of the map.}
\label{fig:denmap}
\end{figure*}  
\subsection{Local Surface Density} \label{kernel}
In order to identify and extract the LSS (filaments \& clusters) in the underlying galaxy distribution, first, we need to estimate the surface density field within the redshift slice defined by our narrow-band H$\alpha$ survey (0.83$\leq$z$\leq$0.86). The surface density field is evaluated using the weighted adaptive kernel density estimator described in detail in Darvish et al. 2014 (in preparation). In this method, we first associate a weight to each galaxy, $i$, in the control sample (consisting of both quiescent \& star-forming galaxies) using its photo-z PDF. The weight for each individual galaxy, $w_i$, is defined as the fraction of its photo-z PDF which lies within the boundaries of the redshift slice. This gives the probability of the galaxy belonging to the redshift slice associated with the H$\alpha$ selected sample. The initial estimate of the surface density associated with the $i$th galaxy, $\hat{\Sigma}_i$, is found by summing over all the weighted fixed kernels placed on the positions of galaxies, $j$, where $i\neq j$:
\begin{equation} \label{eq1}
\hat{\Sigma}_i= \frac{1}{\sum_{\substack{j=1\\j\neq i}}^{N} w_j}\sum_{\substack{j=1\\j\neq i}}^{N} w_j K(\bf r_i,\bf r_j,h)
\end{equation}
where N is the number of galaxies in the control sample, $K(\bf r_i,\bf r_j,h)$ is the fixed kernel, $\bf r_i$ is the position of the galaxy for which the initial estimate of surface density is measured and $\bf r_j$ is the position of the rest of the galaxies. The width of the kernel is expressed by the parameter, $h$, which is a proxy for the degree of smoothing. For the first estimate of the density, this is taken to be fixed. For the kernel smoothing function, we use a 2D symmetric Gaussian defined as:
\begin{equation}
K(\bf r_i,\bf r_j,h)=\frac{1}{2\pi h^2}e^{-\frac{|\bf r_i-\bf r_j|^2}{2h^2}}
\end{equation}
\begin{figure*}
\centering
\includegraphics[width=7.0in]{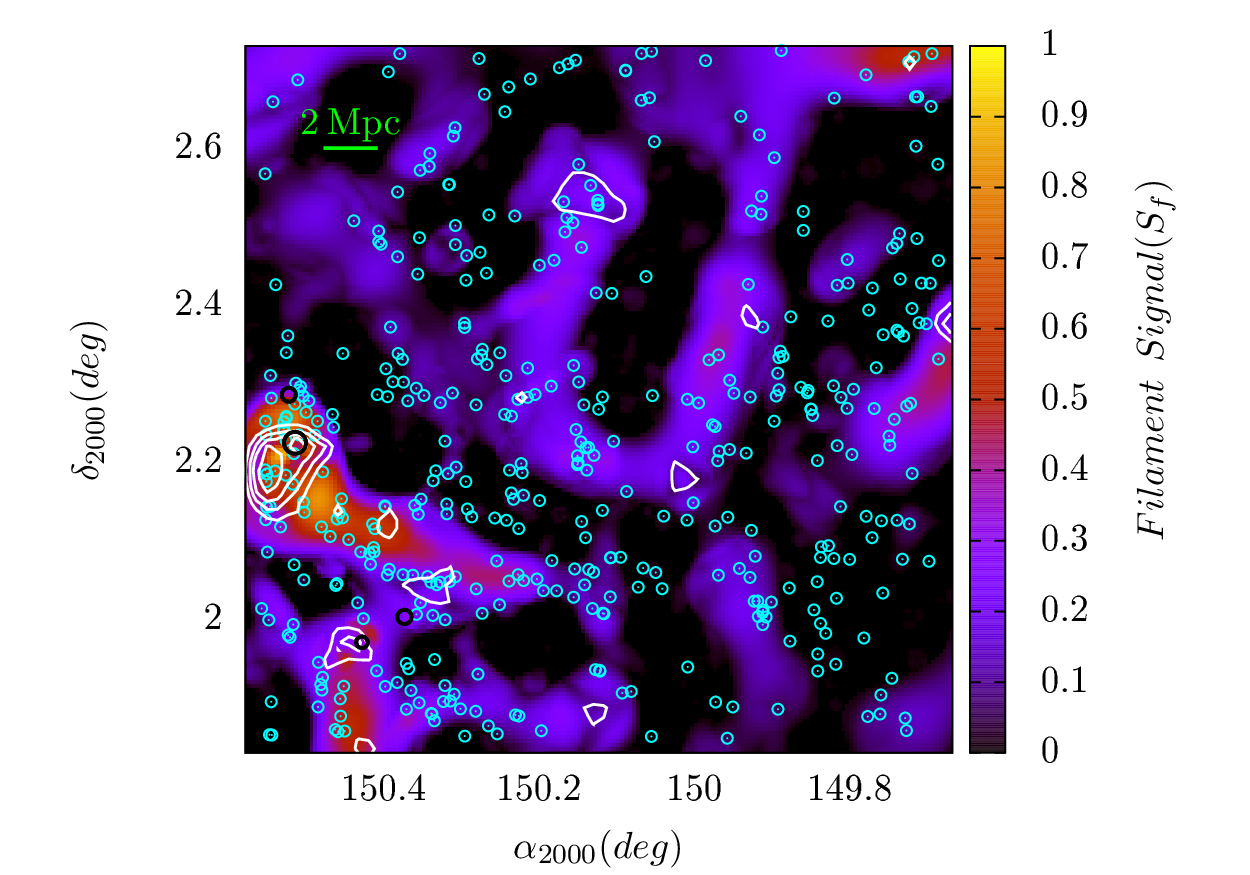}
\caption{ The filament signal map constructed from the Ks-band selected underlying distribution of galaxies using the MMF algorithm for the redshift slice 0.83$\leq$z$\leq$0.86 in the HiZELS area of the COSMOS field. The white contours are extracted from the cluster signal map for the same underlying galaxies as explained in the text. Black circles show the $R_{200}$ region of the X-ray clusters/groups in redshift range 0.82$\leq$z$\leq$0.87 over the same area. Blue circle-dots show the distribution of H$\alpha$ emitters overlaid on the filament signal map. Note the filamentary structure with a great filamentary signal value connecting the over-densities and clusters in the lower left side of this map and its corresponding thread-like distribution of H$\alpha$ emitters tracing the same region.}
\label{fig:filmap}
\end{figure*}
A large kernel width ($h$) results in over-smoothing of the density field which tends to wash out real features while a small value tends to break up regions into smaller uncorrelated substructures. Here, we use a fixed physical length of h=0.5 Mpc which corresponds to the typical value of $R_{200}$ for X-ray clusters \& groups in the COSMOS field \citep{Finoguenov07,George11}. However, a constant value of $h$ for the whole field has the problem that it underestimates the surface density in crowded regions while overestimates in sparsely populated areas. To overcome this problem, we introduce 
adaptive smoothing width, $h_i$,  which is a measure of the local surface density associated with each galaxy, $\hat{\Sigma}_i$. This is defined as $h_i=h\times\lambda_i$, where $\lambda_i$ is a parameter that is inversely proportional to the square root of the surface density associated with the $i$th galaxy, at the position of that galaxy \citep{Silverman86}:
\begin{equation}
\lambda_i\propto\hat{\Sigma}(\bf r_i)^{-0.5} 
\end{equation}
Having the adaptive kernel, we now calculate the surface density field, $\Sigma({\bf r})$, on each location on a fine 2D grid, ${\bf r}$=(x,y) as:
\begin{equation} \label{eq4}
\Sigma(\bf r)= \frac{1}{\sum_{i=1}^{N} w_i}\sum_{i=1}^{N} w_i K(\bf r,\bf r_i,h_i)
\end{equation}
The surface density field is evaluated on a 300x240 grid with a grid size (resolution) of 0.005 deg (corresponding to $\sim$0.125 Mpc physical size at z$\sim$0.85). The estimated surface density field needs to be corrected for incompleteness and contamination in the control sample, since the incompleteness tends to underestimate the density values whereas the contamination has a tendency to overestimate them. Thus, we introduce a correction factor, $\eta$, simply defined as:
\begin{equation}
\eta=\frac{(1-contamination)}{completeness}
\end{equation}
The surface density field corrected for completeness and contamination, $\Sigma_c(\bf r)$, becomes:
\begin{equation}
\Sigma_c(\bf r)=\eta \Sigma(\bf r)
\end{equation}
The density field corresponding to the underlying galaxies near the edge of the field covering H$\alpha$ emitters is biased toward the smaller values, affecting the surface densities near the edge of the field. To minimize this edge effect, we make use of a larger underlying galaxy sample distributed well beyond the H$\alpha$ coverage to estimate densities. Later, we limit our analysis to the area of H$\alpha$ emitters.\\
We stress that since we use galaxy weights, $w_i$s, to estimate the surface density field (equations \ref{eq1} \& \ref{eq4}), our method is immune to spurious structures due to possible random clustering of foreground and background galaxies. These contaminating galaxies would have smaller weights (because only the tail of their photo-z PDF would overlap the redshift slice associated with the H$\alpha$ sample) and therefore, they contribute only marginally to the estimated density field.\\    
We confirm the robustness of the estimated density field in this section by comparing it with the X-ray clusters/groups \citep{Finoguenov07,George11} and voronoi tessellation-based density field of \citealp{Scoville13}. This is presented in Figure \ref{fig:denmap} where we find a very good agreement between our weighted adaptive kernel estimator and the independent estimation based on voronoi tessellation \citep{Scoville13}. There is also a good consistency between the overdense regions and the position of X-ray clusters/groups.

\subsection{Filament \& Cluster Extraction} \label{MMF}
To identify and extract spatial structures such as filaments and clusters in the underlying galaxy distribution, we utilize the 2D version of the Multi-scale Morphology Filter (MMF) algorithm developed by \citealp{Aragon-Calvo07}. This algorithm is able to assign a filament and a cluster signal to each point in the surface density field based on the local geometry of the point in question. If the local geometry of a point in the density field has a greater resemblance to a thread-like structure (filament) than a circular-like feature (cluster), its filamentary signal surpasses its cluster signal. The local geometry of each point is calculated based on the signs and ratio of eigenvalues of the Hessian matrix $H(\bf r)$ which is the second-order derivative of the surface density field:
\begin{equation}
H(\bf r)=
\begin{bmatrix}
\nabla_{xx}\Sigma_c(\bf r)& \nabla_{xy}\Sigma_c(\bf r)\\
\nabla_{yx}\Sigma_c(\bf r)& \nabla_{yy}\Sigma_c(\bf r)
\end{bmatrix}
\end{equation}
Where $\nabla_{ij}$s denote the second derivatives in the i and j directions. Since structures have a variety of physical sizes, this algorithm builds a scale-independent structure map by smoothing the surface density field over a range of physical scales and eventually selecting the greatest cluster and filament signal among all the various signal values at different physical scales. Here, we use a circularly symmetric Gaussian smoothing function with various physical scales in the range $\sim$0.125-1.5 Mpc. Ultimately, each point in the density field has an associated filament-like value between 0 and 1 (filament signal, $S_{f}$) as well as a cluster-like value between 0 and 1 (cluster signal, $S_{c}$) that are interpolated to the positions of galaxies. $S_{f}$ and $S_{c}$ indicate the degree of resemblance of the local environment of any given point to a filament or cluster, respectively (Figure \ref{fig:filmap}).

\subsection{Filament, Cluster \& Field Selection} \label{selection}
\begin{figure}
\centering
\includegraphics[width=3.7in]{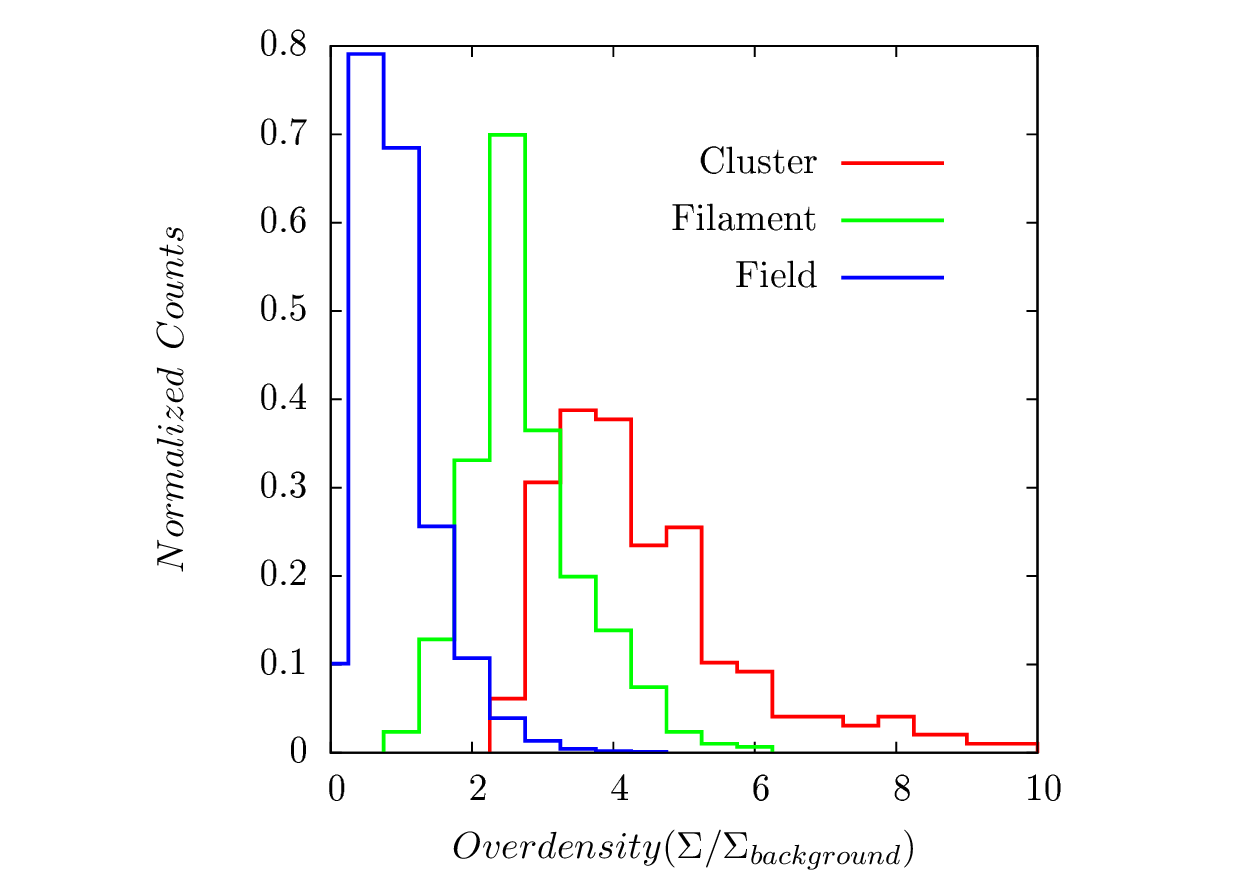}
\caption{Normalized distribution of overdensity values within regions defined as clusters, filaments \& the field. While these structures occupy roughly distinct range of overdensity values, there is significant overlap. We find that intermediate densities correspond mostly to filaments (even though only our analysis actually disentangles them from clusters and the field).}
\label{fig:dens-env}
\end{figure}

Using the techniques developed in section \ref{kernel} and \ref{MMF}, three environments are defined: clusters as regions with a cluster signal greater than 0.5 and also greater than or equal to that of a filament (S$_{c}>$0.5 \& S$_{c}\geq$S$_{f}$), filaments as regions with filament signals greater than 0.5 and greater than that of a cluster (S$_{f}>$0.5 \& S$_{f}>$S$_{c}$) and the remaining regions with both cluster and filament signals less than or equal to 0.5 (S$_{c}\leq$0.5 \& S$_{f}\leq$0.5) as part of the general field. The structures identified over the area covered by the H$\alpha$ narrow-band survey are shown in Figure \ref{fig:filmap}. We demonstrate the filament signal values as a heat map whereas the cluster signal values are shown as contours. We have confirmed the resulting structures in Figure \ref{fig:filmap} by comparing them with the structures depicted from X-ray observations. The X-ray groups \& clusters in the redshift range 0.82$\leq$z$\leq$0.87 \citep{Finoguenov07,George11} are overlaid on our predicted structures in Figure \ref{fig:filmap} and show close agreement.\\
Using the defined environments, we estimate the average number density of H$\alpha$ emitters to be $\delta_{H\alpha}$=1.53$\pm$0.29, 0.45$\pm$0.09 \& 0.13$\pm$0.01 arcmin$^{-2}$ in clusters, filaments \& the field, respectively. The errors are estimated assuming Poisson statistics. $\delta_{H\alpha}$ values clearly show that dense, intermediate \& sparse environments are respectively occupied by clusters, filaments \& the field. A better representation of the density in different classes of environment is seen in Figure \ref{fig:dens-env}. Normalized distribution of overdensity values of the surface density field (estimated in section \ref{kernel}) within regions defined as clusters, filaments \& the field is shown in Figure \ref{fig:dens-env}. Although they inhabit relatively distinct values of overdensity, there is substantial degree of overlap which means that a pure density-based definition of environment is not fully adequate to identify different structures. Intermediate densities do correspond mostly to filaments (even though only our analysis actually disentangles them from clusters and the field). Throughout this work, we use the environment defined based on the MMF algorithm as a robust indicator of the cosmic web.\\
We also note that ideally, one wishes to select clusters and filaments with the highest signal value but practically, this leads to a statistically smaller sample size (especially for H$\alpha$ emitters). The selection cut here (0.5) is due to a trade-off between the sample size and the reality (significance) of clusters and filaments. Nevertheless, we fine-tuned around the selection cut and examined some other values (0.3, 0.4 and 0.6) and found no significant change in the results.\\
\section{Results \& Comparison with Other Studies} \label{result}

\subsection{Fraction of Star-forming Galaxies in the Cosmic Web} \label{fraction}
\begin{figure}
\centering
\includegraphics[width=3.5in]{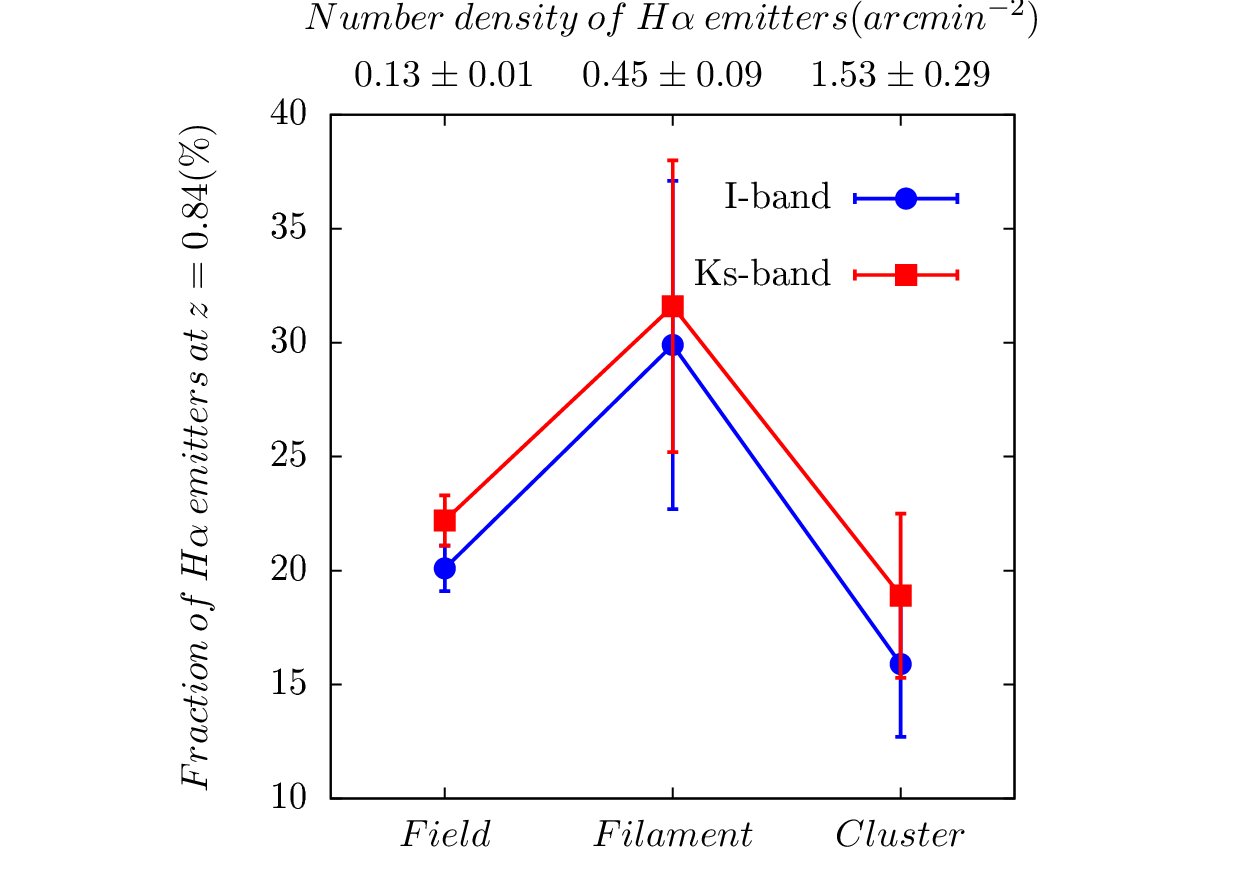}
\caption{ The fraction of H$\alpha$ emitters (the ratio of the number of H$\alpha$ emitters to the total number of underlying galaxies in the control sample) corrected for completeness and contamination in the number of underlying galaxies over different cosmic environments and for both the Ks-band and the I-band selected control samples. The error bars are evaluated assuming Poisson statistics. Note an elevation in the fraction of H$\alpha$ emitters in filaments relative to clusters and the field. This trend is not much affected by photometric selection of the control sample. We also show the average number density of H$\alpha$ emitters (in unit of arcmin$^{-2}$) in clusters, filaments \& the field on top of the figure.}
\label{fig:fraction}
\end{figure}

Using the structures identified in section \ref{selection}, we now study the fraction of star-forming H$\alpha$ emitters (the ratio of the number of H$\alpha$ emitters to the total number of underlying galaxies in the control sample) in filaments, clusters and the field. This fraction is corrected for contamination and incompleteness. In order to examine the effect of photometric selection on our result, we perform this on both our original Ks-band and I-band selected control samples in the COSMOS (the I-band selected catalog is from \citealp{Capak07}). The correction factor, $\eta$, is 0.49 and 0.47 for the Ks and I-band samples, respectively. An increase in the fraction of H$\alpha$ emitters is seen in the filaments relative to both the cluster and the field environments (Figure \ref{fig:fraction}). This trend is not influenced by the photometric selection wavelength (I-band vs. Ks-band) in the underlying galaxy population. The error bars in Figure \ref{fig:fraction} are evaluated assuming Poisson statistics. As noted by \citealp{Ilbert13}, the longer wavelength selection (Ks-band here) can more reliably detect galaxies at intermediate to high redshifts and we mention that the results based on the Ks-band is likely more robust.\\
The enhancement in the fraction of H$\alpha$ emitters in filaments (i.e, intermediate densities) is consistent with previous studies, depicting the importance of intermediate environments; i.e., galaxy groups, outskirts of galaxy clusters and filaments, in the evolution of galaxies. These results hold at low-z (z$\lesssim$0.5, \citealp{Fadda08,Tran09,Biviano11,Geach11,Koyama11,Mahajan12}), intermediate-z (z$\sim$1, \citealp{Koyama08,Koyama10,Sobral11,Pintos-Castro13}) and high-z (z$\gtrsim$1.5, \citealp{Koyama14,Santos14}).\\ 
At z$\sim$1, the closest work to our study in terms of sample selection is that of \citealp{Sobral11} who showed that at z$\sim$0.85, the fraction of H$\alpha$ emitters from both COSMOS- \& UDS-HiZELS surveys, when compared to the underlying population of galaxies at the same redshift, peaks at intermediate environments, in agreement with our result. \citealp{Pintos-Castro13} performed a thorough IR study of a super-structure at z~$\sim$0.85 and showed that although the average star formation does not show any clear environmental dependence (cf. section \ref{distrib}), the fraction of star-forming galaxies indicates that intermediate densities are preferred. They showed that the enhancement in the fraction of star-forming far-IR emitters in medium densities is seen for both optically red and blue far-IR emitters and also in different mass-binned samples. This is also consistent with our results here regarding the enhancement in the fraction of H$\alpha$ emitters in filaments, although they used far-IR emitters as indicators of star-forming galaxies and number density of galaxies as a measure of environment. A combined narrow-band H$\alpha$ and AKARI mid-IR imaging survey of a z=0.81 super-structure by \citealp{Koyama10} revealed that the red H$\alpha$ emitters and mid-IR galaxies are more commonly found in medium density environments such as cluster outskirts, groups and filaments. This reinforced the results of a similar study of the same structure by \citealp{Koyama08}, which showed that the fraction of both red and blue 15-$\mu$m-selected galaxies is highest in medium density environments. Both of these studies by \citealp{Koyama08} \& \citealp{Koyama10} at z$\sim$0.8 agree with our results in this section but we note that we are explicitly identifying filamentary structures as the drivers of such enhancement.\\
At lower redshift (z$\lesssim$0.5), there are several studies consistent with our results. For example, \citealp{Geach11} analyzed a super-structure at z=0.55, comprising several dense regions connected by a filamentary structure, using GALEX NUV and Spitzer MIPS 24$\mu$m data and showed that although on average, there is no significant difference between the SFRs of star-forming galaxies in the super-structure and the randomly selected field (cf. section \ref{distrib}), there are intermediate-scale regions within the super-structure where the probability of a galaxy undergoing star formation is enhanced. This increased probability is analogous to our results here regarding the elevation in the fraction of star-forming galaxies in filaments. At z$\sim$0.2, \citealp{Fadda08} discovered two galaxy filaments associated with a cluster (Abell 1763) and found that starburst galaxies preferentially inhabit the two filaments compared to the cluster core and the outer regions excluding filaments. A more recent study of the same supercluster by \citealp{Biviano11} confirmed the previous results, showing that the filament contains the highest fraction of IR-emitting galaxies. These agree with our results although they used a different prescription to identify filaments and a redshift different from our sample. Studies on galaxy group scales at z$\sim$0.4 by \citealp{Tran09} \& \citealp{Koyama11} also show an overall good agreement with our results in filaments. Recently, \citealp{Stroe14} found that the LF of H$\alpha$ galaxies around the shock-induced radio relics of a cluster at z$\sim$0.2 is significantly boosted compared to the field. We argue that such enhancement could be driven by collapsing filamentary structures when two very massive clusters merge.\\
The enhancement in the fraction of star-forming galaxies at intermediate environments is also seen at higher redshifts. For example, \citealp{Koyama14} showed that for a rich cluster at z$\sim$1.5, H$\alpha$-emitting galaxies are preferentially located in the cluster outskirts and \citealp{Santos14} found huge excess of star-forming fraction relative to the corresponding passive fraction in 1-3 Mpc distance from the center of a massive cluster compared to the cluster core and the field. \citealp{Smail14} also found no significant variation in mean SFR with environment for a structure at z$\sim$1.6---consistent with our results in section \ref{distrib}--- and that the highest density regions of this structure are devoid of the most actively star-forming galaxies (far-IR, submm, MIPS \& radio sources), which instead are preferentially found in intermediate density environments.\\
In our study, we used a robust quantitative technique to identify structures with a special emphasis on filaments as locations with intermediate environments. Although the techniques used to identify the structures are different, our results are in close qualitative agreement with previous studies at different redshifts that were discussed above. We conclude that star-forming galaxies are more commonly found in intermediate density regimes at least out to z$\sim$1.5, regardless of how they are selected (H$\alpha$, IR, etc.) or how their environments are defined. For this study, it seems that filaments are likely responsible for this enhancement.\\
The boost in the star formation activity at intermediate environments is not confined to only an elevation in the star-forming fraction. For example, some studies have shown that the rise in the star formation activity at medium environments is a result of an increase in the mean SFR of galaxies \citep[e.g.][]{Coppin12,Porter07,Porter08} or due to an enhancement in both SFR of galaxies and their fraction at intermediate density regimes \citep[e.g.][]{Fadda08,Sobral11}. Here, we showed that filaments lead to an observed enhancement in the star formation fraction. In the following section, we will investigate whether this enhancement is a consequence of an increase in the SFR of many galaxies in filaments or merely due to some galaxies having their star formation switched on by filaments.    

\subsection{The SFR \& Stellar Mass in the Cosmic Web} \label{distrib}
We now investigate the \textit{observed} distribution of SFR, stellar mass and sSFR in different environments. For the star-forming galaxies in the H$\alpha$ sample, the median SFR, stellar mass and sSFR are independent of the environment (Table 1). The same is true for the underlying star-forming galaxies in the control sample (Table 2). The uncertainties in the median values are estimated from 10000 bootstrap trials.\\
\begin{table}
\begin{center}
{\scriptsize
{{Table 1: Median SFR (based on H$\alpha$ line), Stellar Mass and sSFR in the cosmic web at \lowercase{$z$}$\sim$0.8-0.9 for the H$\alpha$ sample. The errors are estimated by bootstrap resampling from 10000 trials.}} 
\begin{tabular}{lcccc}
\hline
\noalign{\smallskip}
Cosmic Structure & SFR & Stellar Mass & sSFR\\
                 & $\Msolyr$ & log($\Msol$) & log(yr$^{-1}$)\\
\hline
Cluster & 4.47$\pm$1.15 & 10.20$\pm$0.24 & -9.45$\pm$0.22\\
Filament & 4.50$\pm$0.73 & 9.85$\pm$0.10 & -9.32$\pm$0.13\\ 
Field & 3.58$\pm$0.14 & 10.01$\pm$0.06 & -9.43$\pm$0.04\\ 
\hline
\end{tabular}
\label{table:Median_Ha}
}
\end{center}
\end{table}

\begin{table}
\begin{center}
{\scriptsize
{{Table 2: Median SFR (form UV flux), Stellar Mass and sSFR in the cosmic web at \lowercase{$z$}$\sim$0.8-0.9 for the star-forming control sample. The errors are estimated by bootstrap resampling from 10000 trials.}}
\begin{tabular}{lcccc}
\hline
\noalign{\smallskip}
Cosmic Structure & SFR & Stellar Mass & sSFR\\
                 & $\Msolyr$ & log($\Msol$) & log(yr$^{-1}$)\\
\hline
Cluster & 9.57$\pm$1.41 & 9.76$\pm$0.04 & -8.81$\pm$0.05\\
Filament & 11.13$\pm$1.62 & 9.79$\pm$0.06 & -8.64$\pm$0.05\\ 
Field & 8.97$\pm$0.30 & 9.68$\pm$0.01 & -8.67$\pm$0.02\\ 
\hline
\end{tabular}
\label{table:Median_control}
}
\end{center}
\end{table}
\begin{figure}
\centering
\includegraphics[width=3.7in]{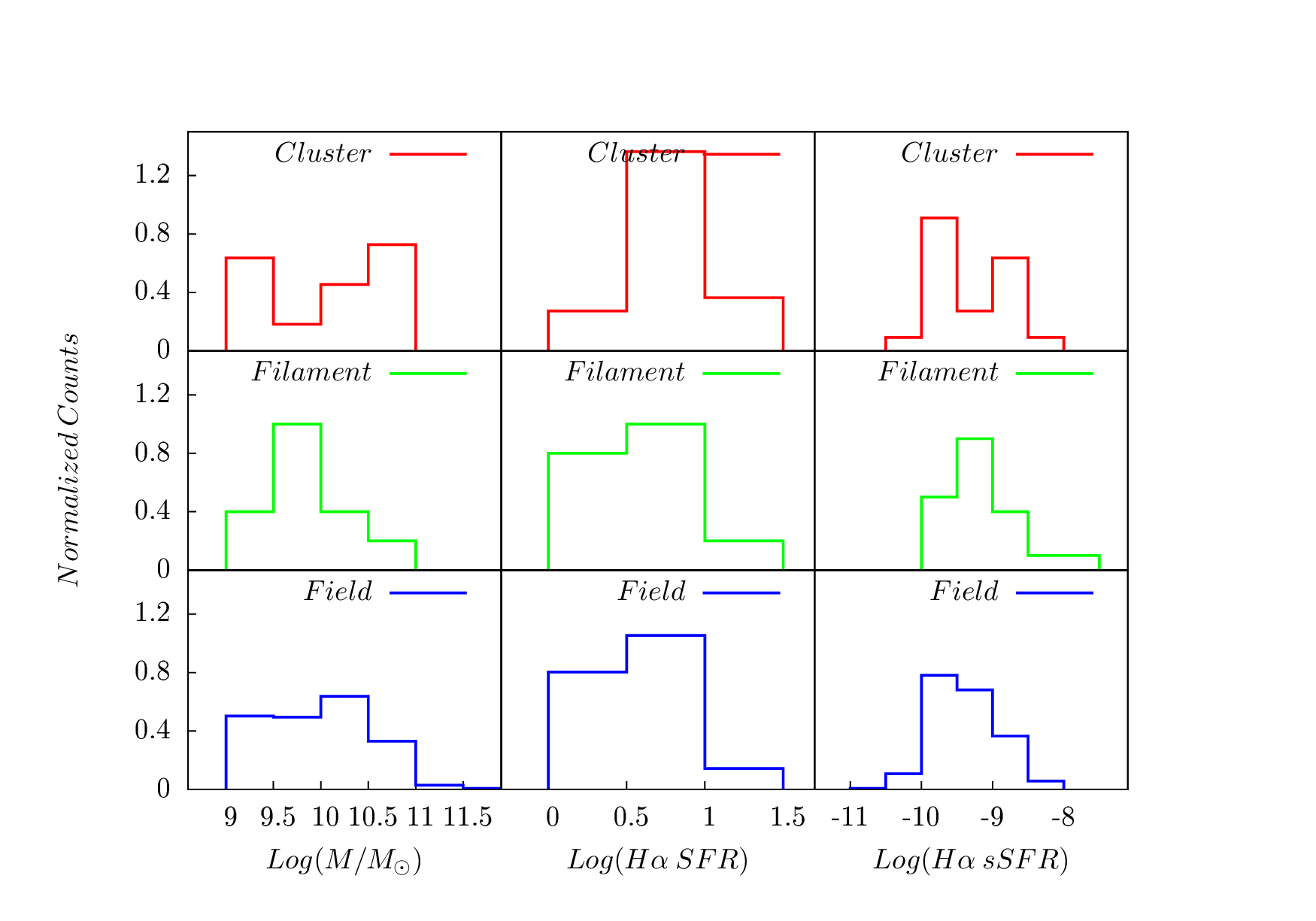}
\caption{Normalized stellar mass, SFR and sSFR distribution of H$\alpha$ emitters in clusters, filaments and the field at z$\sim$0.8-0.9. These observed distributions are similar in different environments. Using the K-S test we show that it is statistically unlikely that the above-mentioned distributions are drawn from different parent populations. The K-S test p-value is $>$ 0.01 in all cases.}
\label{fig:dist_Ha}
\end{figure}
\begin{figure}
 \centering
  \includegraphics[width=3.7in]{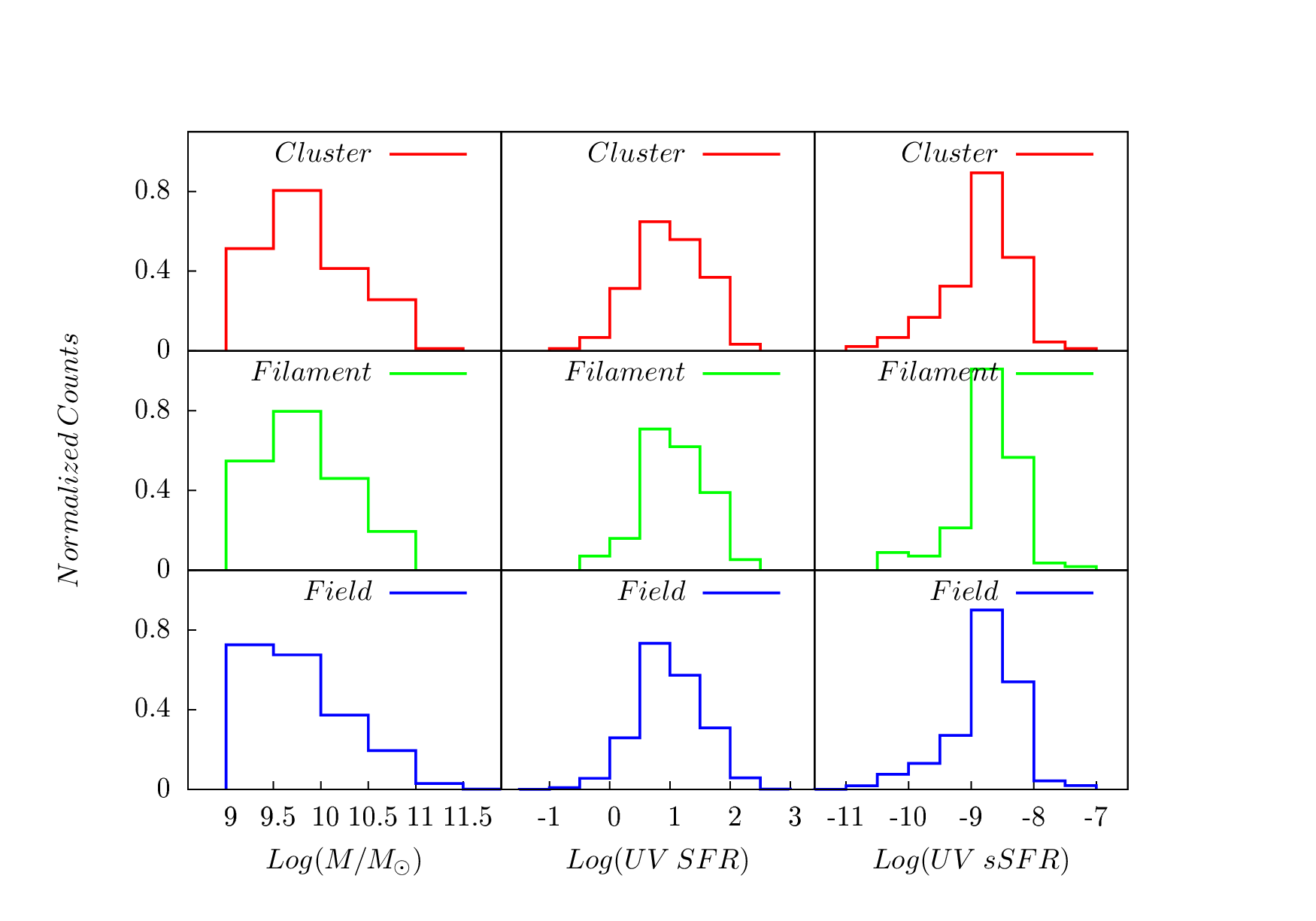}
\caption{Same as Figure \ref{fig:dist_Ha} but for star-forming galaxies in the control sample.}
\label{fig:dist_control}
\end{figure}
A comparison between the observed distribution of H$\alpha$-based SFRs in different environments reveals similarities (Figure \ref{fig:dist_Ha}), showing that the environment does not significantly affect the SFR of selected star-forming galaxies. The observed distribution of the stellar mass and the sSFR is also found to be independent of the environment for the H$\alpha$ sample (Figure \ref{fig:dist_Ha}). These results also hold true for the selected star-forming galaxies in the control sample (Figure \ref{fig:dist_control}).\\
\begin{table*}
\begin{center}
{\scriptsize
{{Table 3: Results of the K-S test for distributions of H$\alpha$ SFR, Stellar Mass and sSFR from the H$\alpha$ sample in different cosmic structures. The p-values are the probabilities that the two distributions are drawn from the same parent population.}}
\begin{tabular}{lcccc}
\hline
\noalign{\smallskip}
Cosmic Structures & p-value & p-value & p-value\\
                  & (logSFR distributions) & (logMass distributions) & (logsSFR distributions)\\
\hline
Cluster \& Filament & 0.22 & 0.05 & 0.22\\
Cluster \& Field & 0.02 & 0.08 & 0.44\\
Filament \& Field & 0.14 & 0.24 & 0.18\\                
\hline
\end{tabular}
\label{table:K-S_Ha}
}
\end{center}
\end{table*}

\begin{figure}
\centering
\includegraphics[width=3.5in]{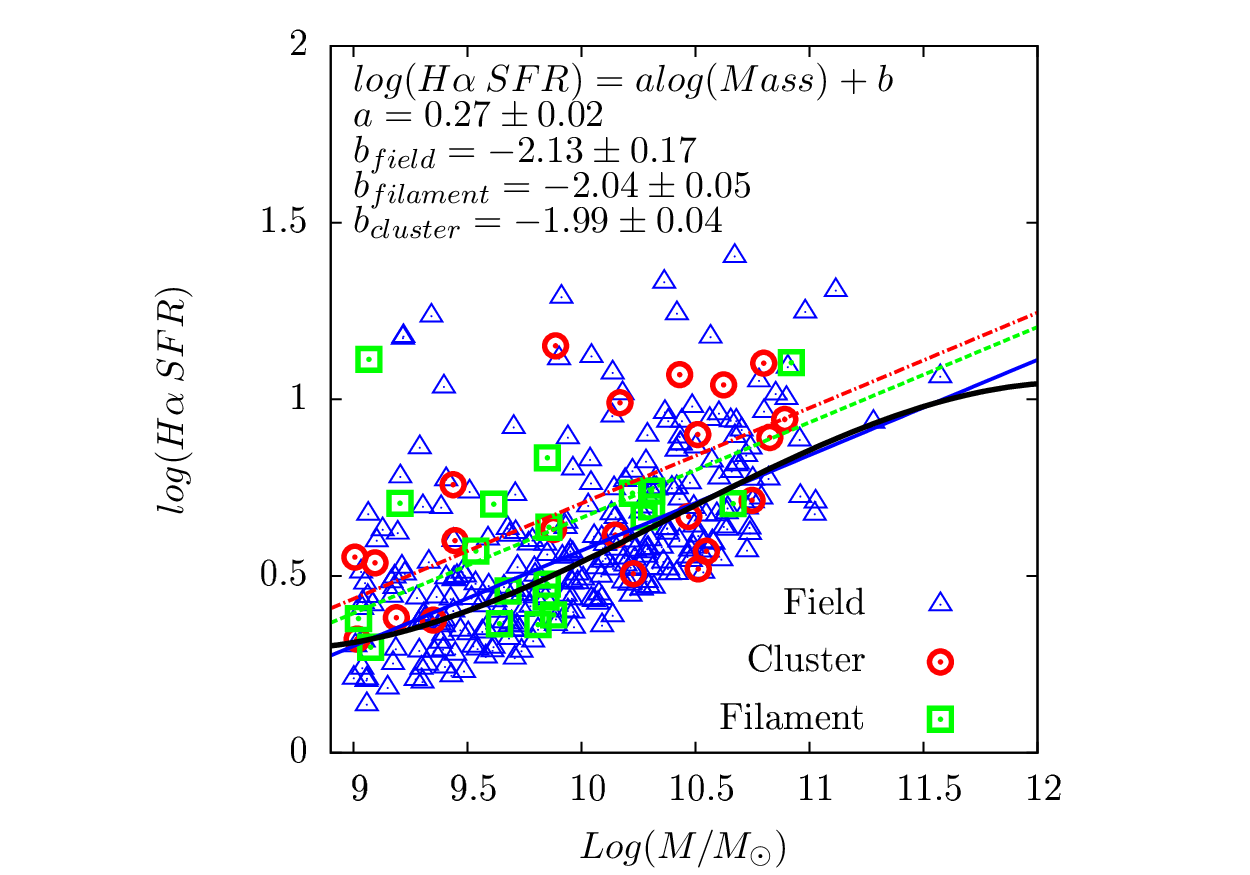}
\caption{The SFR-Mass relation for H$\alpha$ emitters in clusters, filaments and the field. The best-fit lines for clusters, filaments and the field data are displayed with dot-dash (red), dashed (green) and solid (blue) lines, respectively. The results of the best-fit parameters are given on the top left corner of the figure. The observed SFR-Mass relation does not depend on the environment. A careful 2D K-S test also shows that it is statistically unlikely that the 2D log(SFR)-log(Mass) distributions in clusters, filaments and field are drawn from different parent populations (p-value $>$ 0.01). The black solid line also shows the H$\alpha$ selection limit. Note the large envelope of galaxies near the selection limit.} 
\label{fig:MS_Ha}
\end{figure}

\begin{figure}
\centering
\includegraphics[width=3.5in]{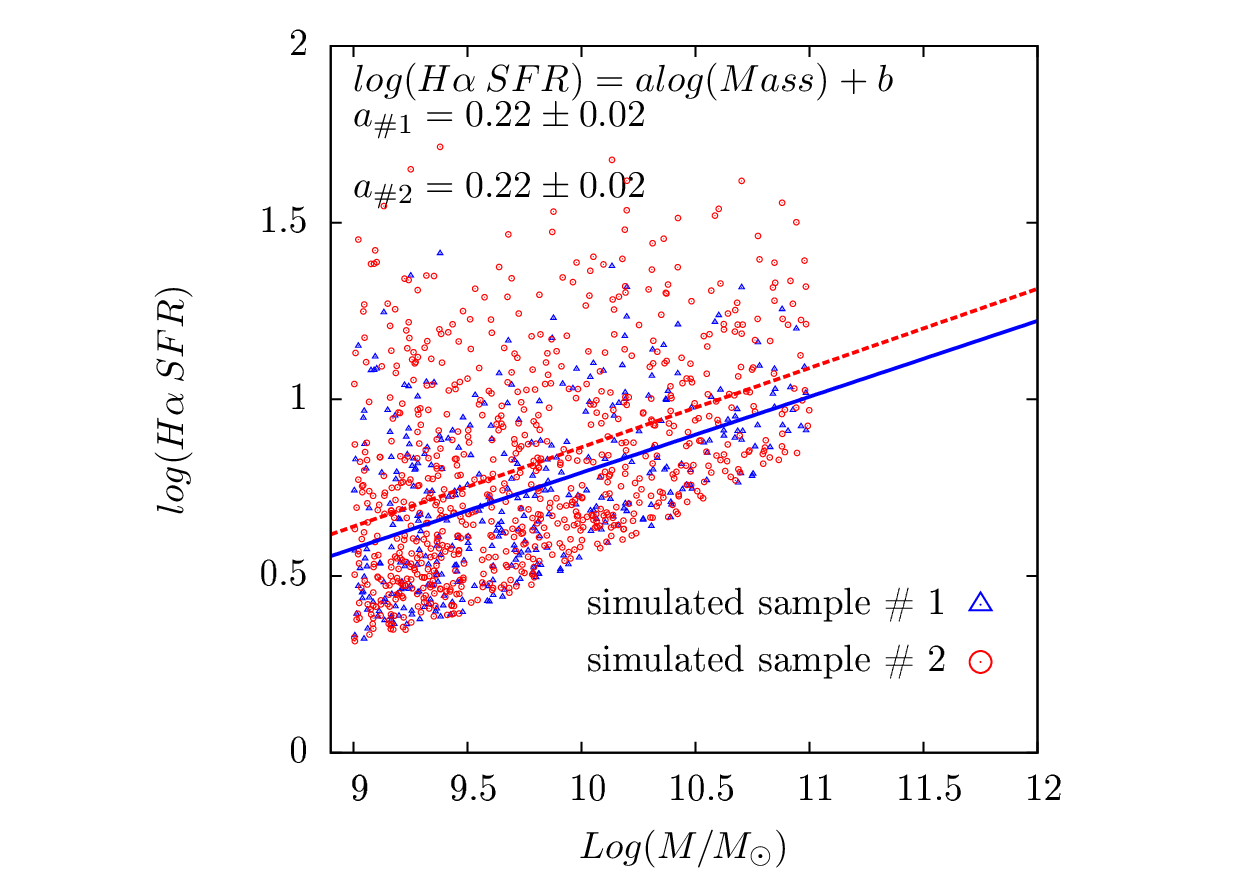}
\caption{ The H$\alpha$ SFR-Mass relation for two sets of simulations. These simulations are subjected to the observational selection biases of the H$\alpha$ star-forming sample. The intrinsic SFR of galaxies for simulation \# 2 (red symbols) is twice that of simulation \# 1 (blue symbols). Despite the intrinsic difference between the SFRs of galaxies in these simulations, their SFR-Mass relation is almost the same within statistical uncertainties.}
\label{fig:MS_sim}
\end{figure}

\begin{figure}
\centering
\includegraphics[width=3.5in]{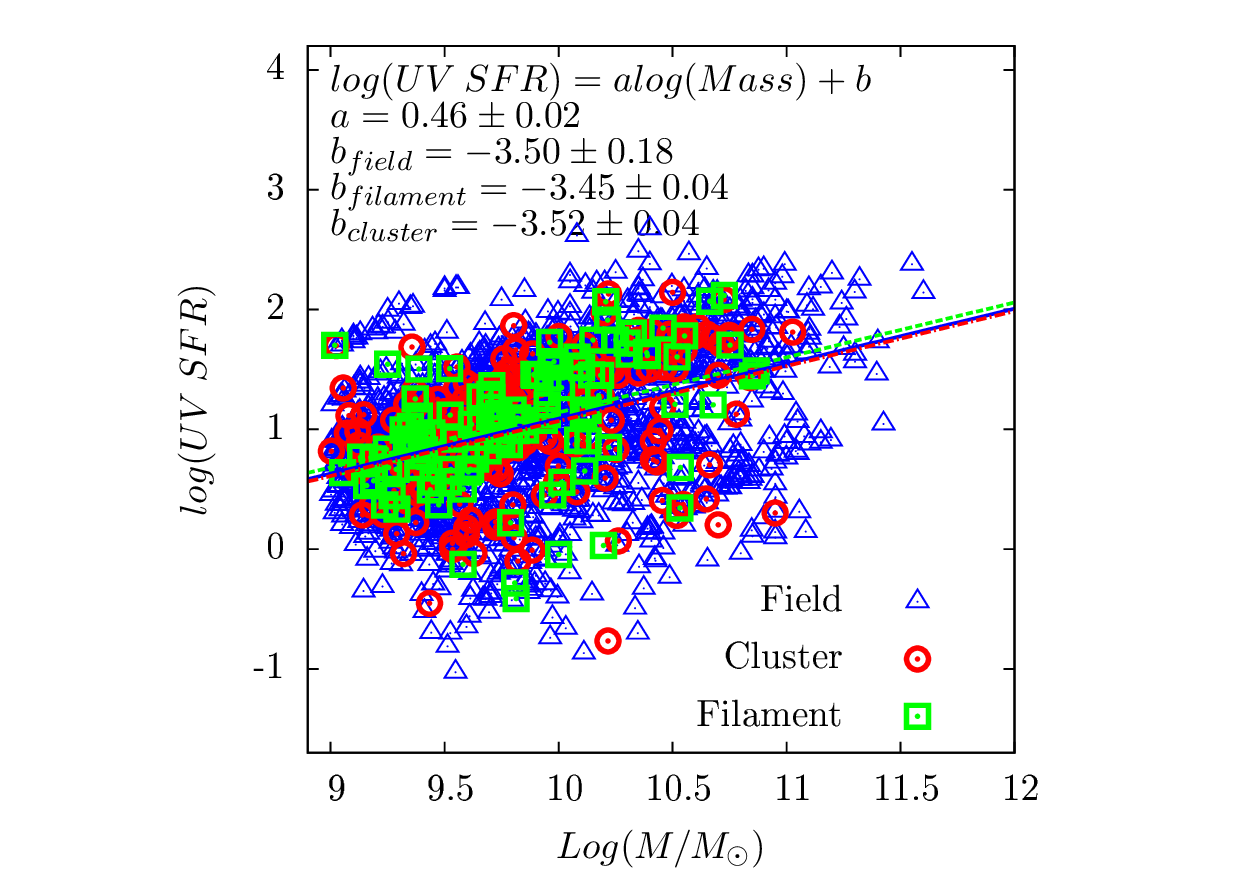} 
\caption{Same as Figure \ref{fig:MS_Ha} but for star-forming galaxies in the control sample.}
\label{fig:MS_control}
\end{figure}

\begin{table*}
\begin{center}
{\scriptsize
{{Table 4: Results of the K-S test for distributions of UV SFR, Stellar Mass and sSFR from the star-forming galaxies in the control sample in different cosmic structures. The p-values are the probabilities that the two distributions are drawn from the same parent population.}}
\begin{tabular}{lcccc}
\hline
\noalign{\smallskip}
Cosmic Structures & p-value & p-value & p-value\\
                  & (logSFR distributions) & (logMass distributions) & (logsSFR distributions)\\
\hline
Cluster \& Filament & 0.62 & 0.46 & 0.09\\
Cluster \& Field & 0.66 & 0.02 & 0.03\\
Filament \& Field & 0.19 & 0.12 & 0.44\\                
\hline
\end{tabular}
\label{table:K-S_control}
}
\end{center}
\end{table*}
We perform K-S tests to compare distribution of SFRs, stellar Masses and sSFRs in different environments and the results are tabulated in tables 3 \& 4 for both the H$\alpha$ emitters and the star-forming galaxies in the control sample. Here, the p-value is the probability that the two considered distributions are drawn from the same underlying parent population. It is unlikely that the observed distributions of SFRs, stellar masses and sSFRs in different environments are extracted from different parent populations, with the evaluated p-value $>$ 0.01 in all cases.\\
The results in this section are consistent with other independent studies at different redshifts \citep{Peng10,Geach11,Wijesinghe12,Koyama10,Koyama13a,Koyama13b,Koyama14, Feruglio10,Ideue12,Muzzin12,Pintos-Castro13,Bouche05,Grutzbauch11,Brodwin13,Hayashi14,Santos14,Smail14}. 
Here, we found no significant evidence that the cosmic web affects the stellar mass, SFR \& sSFR of the \textit{observed} star-forming galaxies at z$\sim$1; a result that might be partially explained by selection biases.\\    
To further investigate this, we study the environmental (filament, cluster and field) dependence of the SFR-Mass relation---often known as the ``Main-Sequence" of star-forming galaxies---for the H$\alpha$ emitting galaxies and the star-forming galaxies in the control sample. 
The mean SFR-Mass relation for the H$\alpha$ sample appears to be independent of the environment (Figure \ref{fig:MS_Ha}). Using the nonlinear least-squares Marquardt-Levenberg algorithm, we perform a linear fit (log(SFR)=alog(Mass)+b) to the log(SFR)-log(Mass) relation for H$\alpha$ emitters in filaments, clusters and the field, separately. For clusters and filaments, the slope of the linear fit is fixed to the slope of the best-fitting line for the field H$\alpha$ sample. The slope of the best-fitting line is a=0.27$\pm$0.02 with the intercept being b=-1.99$\pm$0.04, -2.04$\pm$0.05 and -2.13$\pm$0.17 for cluster, filament and the field H$\alpha$ emitters, respectively. The intercepts are consistent within the uncertainties. We also evaluate the observed scatter around the mean SFR-Mass relation in different environments. The observed scatter for the H$\alpha$ sample is found to be $\lesssim$0.2 dex, independent of the environment. It is worth mentioning that the typical observational uncertainty in the H$\alpha$ SFR (section \ref{M-SFR-Ha}) is greater than the observed scatter around the mean SFR-Mass relation. It is noteworthy that a slight increase in the median H$\alpha$ SFR is seen in clusters compared to the field. However, this enhancement is not significant and is within uncertainties.\\
In another attempt to quantify the environmental dependence of the SFR-Mass relation, we use the generalized two-dimensional K-S test \citep{Fasano87} to determine the probability that the populations of cluster, filament and field H$\alpha$ emitters on the log(SFR)-log(Mass) plane are drawn from the same parent distribution. The derived p-value is 0.09 (cluster vs. filament), 0.06 (cluster vs. field) and 0.12 (filament vs. field), implying that they are unlikely to be drawn from different parent populations on the logSFR-logMass plane (The p-value is $>$ 0.01 in all cases).\\
The environmental independence of the median SFR and the mean SFR-Mass relation for the H$\alpha$ star-forming galaxies might be effected by the selection biases in the data. In other words, the median SFR of star-forming galaxies might be intrinsically different in different environments but due to selection effects, this difference does not mirror itself into the observed median SFR and/or the mean SFR-Mass relation.\\
Since the H$\alpha$ selected sample is flux limited (i.e., SFR limited), the SFR-Mass relation could be entirely driven by the fixed H$\alpha$ SFR limit, the shape of the H$\alpha$ luminosity function and the mass-dependent dust correction we use in this work. Due to the shape of the luminosity function and the SFR limit of the H$\alpha$ sample ($\sim$ 1.5 $\Msolyr$), the majority of galaxies tend to crowd near the selection limit which defines a relatively narrow horizontal strip in the SFR-Mass plane. A mass-dependent dust correction to the H$\alpha$ sample then introduces a slope in the relation. Since these selection biases apply equally to all environments, this leads all environments to give the same results.\\
In order to check this, we perform a simulation. We select a mock sample of galaxies with the stellar mass randomly drawn from 9$<$log(M/$\Msol$)$<$11 and their SFR randomly drawn from the H$\alpha$ luminosity function based on the HiZELS survey (L$^{*}$=42.25 ergs$^{-1}$, $\alpha$=-1.56). We emphasize that in randomly selecting these galaxies, no assumption about an intrinsic correlation between SFR and stellar mass is made. We dust-redden the sample using the mass-dependent dust correction equation used in this work (section \ref{M-SFR}) and apply the HiZELS H$\alpha$ SFR limit (1.5 $\Msolyr$) to mimic the observational selection of the HiZELS survey. We later undo the dust extinction on the galaxies that passed the SFR selection limit. Figure \ref{fig:MS_sim} (blue symbols) shows the SFR-Mass relation for these simulated galaxies considering the selection effects. This resembles what we observe in Figure \ref{fig:MS_Ha}, despite the fact that we initially introduced no intrinsic relation between SFR and stellar mass \textit{a priori}. The slope of the SFR-Mass relation for these simulated galaxies is a=0.22$\pm$0.02, similar to that of Figure \ref{fig:MS_Ha}, within errors.\\
To further investigate the selection effects on the results, we double the intrinsic SFR of every galaxy and assume that they happen to be located in filaments. This means that more galaxies will pass the H$\alpha$ selection limit due to the shape of the luminosity function. The fraction of galaxies that now pass the selection limit is increased by a factor of $\sim$2.2 (cf. Figure \ref{fig:fraction}). However, the median SFR and the mean SFR-Mass relation do not change significantly. Figure \ref{fig:MS_sim} (red symbols) shows the SFR-Mass relation for the new simulated galaxies (in filaments) with twice the intrinsic SFR. This SFR-Mass relation looks similar to that of blue symbols although the intrinsic SFR of each individual galaxy is doubled. The median SFR for detected galaxies is only slightly increased form 5.30$\pm$0.22 to 5.91$\pm$0.19. The slope and intercept values are now a=0.22$\pm$0.02 and b=-1.38$\pm$0.17, respectively, remain almost the same within uncertainties (these values are a=0.22$\pm$0.02 and b=-1.36$\pm$0.20 for blue symbols).\\
We conclude that the environmental invariance of the median SFR and the mean SFR-Mass relation for the H$\alpha$ star-forming galaxies is partly due to selection biases. In other words, the intrinsic SFR of star-forming galaxies might be a function of the environment but is not completely seen in the observed SFR-Mass relation or the median SFR, due to selection effects. The intrinsic enhancement of SFR of H$\alpha$ emitters in filaments, together with the selection biases, has the same effect of elevating the observed fraction of H$\alpha$ star-forming galaxies in filaments (results in section \ref{fraction}), while leaving the observed median SFR and the mean SFR-Mass relation almost intact. Therefore, from the H$\alpha$ sample alone, the selection biases are sufficiently strong that we cannot conclude whether it is the average SFR of all galaxies that increases in filaments, or whether some galaxies have their star formation switched on by filaments that lead to an enhancement in the star-forming fraction.\\
We find similar results for the SFR-Mass relation based on the rest-frame NUV-$r^{+}$ versus $r^{+}$-J selected star-forming galaxies in the control sample (Figure \ref{fig:MS_control}); i.e., the mean SFR-Mass relation for selected star-forming galaxies in the control sample is found to be independent of the environment. As explained in section \ref{M-SFR-control}, the SFR of the control sample is based on the UV continuum and IR flux (where available). A linear fit to the logSFR-LogMass data results in a slope of a=0.46$\pm$0.02 and intercepts b=-3.52$\pm$0.04, -3.45$\pm$0.04 and -3.50$\pm$0.18 for cluster, filament and field star-forming galaxies in the control sample, respectively (the slope is fixed to that for the field sample). We evaluate the intrinsic scatter around the mean SFR-Mass relation in different environments, after subtracting in quadrature the typical observational uncertainty in SFR (section \ref{M-SFR-control}) from the observed scatter. The intrinsic scatter for the star-forming galaxies in the control sample is $\sim$0.4 dex, regardless of the environment. Also, similar to the results from the H$\alpha$ sample, the 2D K-S test exhibits that the environmental discrepancies are insignificant, with p=0.43, 0.05 and 0.11 for cluster \& filament, cluster \& field and filament \& field, respectively.\\
We stress that the estimated SFRs for the H$\alpha$ emitters and the underlying star-forming galaxies have different origins, calculated based on different recipes and are based on different diagnostics. We apply the K-S test to the distribution of log(H$\alpha$ SFR/UV SFR) for the H$\alpha$ emitters with available UV SFR in three different environments. The K-S test p-value is p=0.59, 0.12 and 0.53 for cluster \& filament, cluster \& field and filament \& field, respectively, indicating that the result is insensitive to the SFR estimators used in this analysis. We emphasize that our main purpose here was to investigate whether the environment affects the SF activity in star-forming galaxies and not to compare different SFR diagnostics.\\
Similar to the H$\alpha$ selected sample, the results based on the star-forming galaxies in the control sample are not completely free from selection biases. It is more difficult to model the effect of the selection bias for the star-forming galaxies in the control sample. However, they might be affected by similar potential biases. The selection of these star-forming galaxies is based on some color-color cuts and if their SFRs were to increase, then some galaxies which did not initially pass the color selection criteria would now do so, probably leading to the same bias.\\
The environmental independence of the SFR-Mass relation for the H$\alpha$ sample, as well as the star-forming galaxies in the control sample, is in agreement with previous studies at low-z (z$\lesssim$0.5, \citealp{Peng10,Biviano11,Wijesinghe12,Koyama13a,Lin14}), intermediate-z (z$\sim$1, \citealp{Koyama10,Peng10,Ideue12,Muzzin12,Koyama13a,Lin14}) and high-z (z$\gtrsim$1.5, \citealp{Hayashi11,Grutzbauch11,Koyama13a,Koyama14}).\\
These studies cover a wide redshift range (0.05$\lesssim$z$\lesssim$3) with their galaxy samples selected differently from a variety of surveys (SDSS, COSMOS, zCOSMOS, GAMA, GCLASS, Pan-STARRS1, GOODS-NICMOS \& some specifically targeted structures). They have used different proxies for the definition of environment (the majority of them are based on the local number density of galaxies) and the SFR was estimated with different indicators. This includes dust-corrected H$\alpha$-based SFR \citep{Peng10,Koyama10,Koyama13a,Koyama14}, [OII]$\lambda$3727 \citep{Hayashi11,Ideue12,Muzzin12}, total IR luminosity \citep{Biviano11}, extinction-corrected H$\alpha$ EW \citep{Wijesinghe12}, SED template-fitting \citep{Peng10}, rest-frame UV flux \citep{Grutzbauch11} and rest-frame U \& B magnitudes \citep{Lin14}. They all have shown that for star-forming galaxies, the SFR-Mass or sSFR-Mass relation is almost independent of the environment, consistent with our results.\\
We found that the SFR-Mass relation for star-forming galaxies does not depend on the environment, regardless of the methods used to define environment or to estimate SFR. However, it is important to note that the environmental independence of the SFR-Mass relation, as seen in this work and some similar studies, might be partially due to selection effects. 

\section{Discussion} \label{dis}
We have shown that the fraction of star-forming galaxies varies with environment and is peaked in filamentary structures. We have also shown that the \textit{observed} properties of the star-forming galaxies (such as their SFR-Mass relation) are independent of the cosmic web. However, due to selection biases in the current data, we are unable to discriminate between two scenarios: (1) The observed enhancement in the fraction of star-forming galaxies in filaments is the result of an
elevation in the SFR of star-forming galaxies. (2) there is a real intrinsic enhancement of star-forming galaxies in filaments with respect to clusters and the field.\\
It is possible to disentangle these two scenarios with a significantly deeper (and/or wider) H$\alpha$ sample. This would equip us with a mass-dependent luminosity function and a significantly broader range of intrinsic SFRs in different environments. Indeed, a deeper/wider sample would introduce less selection biases in the results.\\  
The enhancement of star formation activity in filaments compared to denser regions like clusters could be attributed to a milder galaxy-galaxy harassment and interaction (\citealp{Lavery88,Moore96}; see also \citealp{Coppin12}) in the former. The environment of filaments is not hot compared to clusters (typical temperature of filaments is $\sim10^{5}-10^{7}$K; e.g. see \citealp{Cen06,Werner08,Zappacosta02,Nicastro05}). Therefore, galaxies in filaments can still hold their gas content to form stars. Ram-pressure stripping \citep{Gunn72,Abadi99} is practically not effective in suppressing the star formation in filaments compared to clusters, given the fact that (1) the IGM within filaments is less dense, (2) galaxies in filaments, on average, have a smaller velocity dispersion with respect to the background gaseous medium and (3) the time-scale of SF events in galaxies is shorter than the effective time-scale for ram-pressure stripping ($\sim$Gyr). Notice, however, that this argument depends on the galaxy mass. In dwarf galaxies, their shallow potential wells can provide a relatively small restoring force from the ram-pressure force of the IGM in filaments, with the end result of significant gas stripping in low mass galaxies due to the cosmic web at typical gas densities and velocities \citep[e.g.][]{Benitez-Llambay13}. Instead, for galaxies in the mass range studied here, the restoring force of the dark matter halo + baryons can be more than 3 orders of magnitude larger than the inner regions of dwarfs, turning ram-pressure stripping from filaments largely inefficient.\\
Another possibility for the enhancement in the fraction of star-forming galaxies in filaments might be due to a more efficient gas accretion rate in filamentary structures. A more efficient gas accretion rate in filaments is able to increase the availability of cold gas for galaxies which in turn, enhances the intrinsic SFR of galaxies in filaments and would affect their SFR-Mass relation. However, if the selection biases turned out to be a minor issue and the SFR-Mass relation were indeed independent of the environment, the gas accretion rate, as the main driver of the SFR-Mass relation \citep{Dutton10}, would become independent of the environment and could not possibly account for the increased fraction of star-forming galaxies in filaments.\\
As galaxies move along filaments toward central region of galaxy clusters, it is likely that a fraction of them survive their passage through the cluster core and fling back out into less dense filamentary structures. It is unlikely that these backsplash galaxies \citep{Gill05,Pimbblet11} significantly influence our results. Since these systems have already entered the cluster environment, their star formation has been likely truncated due to ram pressure stripping and gravitational tidal field of the cluster, well before they come back to filaments. Moreover, backsplash galaxies become important within $\sim$ 1-2 virial radius of clusters \citep{Pimbblet11,Mamon04}, a region which does not cover a significant portion of the filaments in this study.\\ 
Here, we argue that the higher number density of galaxies in filaments compared to the field and their lower velocity dispersion compared to those in clusters, increase the chance of gravitational interaction between galaxies, which in turn results in a higher chance of star-forming galaxies to be found in filaments (and/or a higher SF activity in filaments). Our results are consistent with a physical process that might be related to mild galaxy-galaxy interactions, as numerical simulations have shown that interactions can trigger star formation due to tides and compression of the gas \citep[e.g.][]{Barnes91,Springel00,Cox08}.
The scenario proposed here could induce several other observational signatures, such as a larger fraction of irregular morphologies, close galaxy pairs and AGNs (interaction-induced) associated to filaments than in clusters/field.\\
To further support this idea, we investigated the morphology of H$\alpha$ sample using the available ZEST catalog \citep{Scarlata07}. Interestingly, we find that the fraction of irregular morphologies is higher in filaments compared to that of clusters and the field. The fraction of irregular H$\alpha$ emitters is 32$\pm$12\%, 8$\pm$5\% and 19$\pm$2\% in filaments, clusters and the field, respectively (see also \citealp{Sobral09} for an overall number of irregulars). However, we mention that the uncertainties in the fraction of irregular H$\alpha$ emitters is relatively large and we need a larger number statistics to robustly constrain it. Another piece of evidence arises from the fraction of galaxy pairs in different environments. We find that the fraction of close galaxy pairs (projected distance $<$20 Kpc) in the control sample is 7.7$\pm$2.2\%, 5.4$\pm$1.4\% and 2.7$\pm$0.3\% in filaments, clusters and the field, respectively, higher in filaments \& clusters compared to the field. We also inspected the fraction of AGNs in different environments but due to very small number statistics, we cannot make any conclusions with the AGNs. We mention that all of the above-mentioned observational signatures can be constrained vigorously with a larger sample, a possibility that we would like to explore in the near future. 

\section{Conclusions} \label{concl}
\begin{itemize}
\item Our results improve previous studies based on intermediate environments and agree quantitatively and qualitatively with them, but show that what is often called ``intermediate" densities and/or ``group" environments are likely to be filaments.
\item Our results show that the observed fraction of star-forming galaxies is enhanced in filaments, relative to the field and cluster environments at z$\sim$0.8-0.9. This enhancement in the fraction of star-forming galaxies in filaments is either intrinsic and/or due to a boost in the SFR of star-forming galaxies in filaments when combined with the selection biases. A possible physical interpretation for this trend is mild galaxy-galaxy interactions.
\item We also find that - in agreement with other studies - the observed SFR-Mass relation and its intrinsic scatter, median SFR, stellar mass and sSFR is mostly independent of environment for selected H$\alpha$ emitters and underlying star-forming galaxies at z$\sim$0.8-0.9. However, we mention that the environmental independence of the median SFR and the SFR-Mass relation for star-forming galaxies is affected by the selection biases; i.e., the intrinsic SFR of star-forming galaxies might vary with environment without being observed in the median values or the SFR-Mass relation, as a result of selection effects. Indeed, if the environmental invariance of SFR-Mass relation turns out to be real (and not much affected by the selection biases), this means that the environment is most relevant to set the star-forming fraction, or to determine if a galaxy is actively forming stars or not (but when it forms stars, on average, it lies on the same relation across environments). This implies that the physical processes involved in triggering/quenching star-formation should act in a relatively short time-scale (assuming the SFR-Mass relation is actually independent of environment and not much affected by selection biases).
\item Since our analysis is limited to a narrow redshift slice ($\Delta$z=0.03, equivalent to a comoving radial length of $\sim$80 Mpc at z$\sim$0.84), we discriminate against filaments oriented radially and we miss possible sub-structures along the line of sight. However, this is a limitation to all other similar photo-z based studies. A better understanding of the geometry of structures can be achieved by 3D reconstruction of them using a large spectroscopic sample. We plan to further investigate this analysis using a large spectroscopic data taken for this structure, combined with the spectra available from the literature. We mention that this analysis can be repeated to look for time evolution by applying it to the whole COSMOS field out to z$\sim$3 (in preparation).
\end{itemize}

\section*{acknowledgments}
We gratefully thank the referee for thoroughly reading the original manuscript and providing very useful comments that improved the quality of this work. DS acknowledges financial support from LKBF, the Netherlands Organisation for Scientific research (NWO) through a Veni fellowship, from FCT through a FCT Investigator Starting Grant, a Start-up Grant (IF/01154/2012/CP0189/CT0010) and the grant PEst-OE/FIS/UI2751/2014. BD would like to thank Miguel Arag{\'o}n-Calvo for his useful comments on an earlier draft.

\bibliographystyle{apj} 
\bibliography{Filament-COSMOS-HiZELS-1}

\end{document}